\documentclass[12pt]{article}%,a4paper
\usepackage{graphicx,bm,epsf,amsmath,amsthm,amscd,amsfonts,amssymb,amsbsy,color}
\usepackage{mathabx} % for the command \widebar{\cdots}, OR \overline{\phantom{|}...}
\usepackage{type1ec}
\usepackage[T1,T2A]{fontenc}      % LCY
\usepackage[cp1251]{inputenc}   % cp1251
\providecommand{\href}[2]{#2}
\usepackage[hyphens,spaces,obeyspaces]{url}%must be first!
\usepackage[colorlinks,breaklinks,allcolors=blue]{hyperref}%[dvips,
\usepackage{doi}
\newcommand{\e}{{\,\rm e}}

\newcommand{\T}{T}

\newcommand{\be}{\begin{equation}}
\newcommand{\ee}{\end{equation}}
\newcommand{\nn}{\nonumber}
\newcommand{\bn}{\bar n}
\newcommand{\bym}{\bar y_{\rm max}}
\newcommand{\bum}{\bar u_{\rm max}}

\newcommand {\mm}[1]{\quad\mbox{#1}\quad}
\newcommand {\MM}[1]{\qquad\mbox{#1}\qquad}
\newcommand{\bc}{\begin{center}}
\newcommand{\ec}{\end{center}}
\newcommand{\bi}{\begin{itemize}}
\newcommand{\ei}{\end{itemize}}
\begin{document}
\title{\bf The cell fluid model with Curie-Weiss interactions: special cases and analytical results}
\author{{\bf O. A. Dobush, M. P. Kozlovskii, I. V. Pylyuk, R. V. Romanik,}\\
{\bf and M. A. Shpot}\\[2mm]
{\it Yukhnovskii Institute for Condensed Matter Physics}\\
{\it of the National Academy of Sciences of Ukraine, 79011 Lviv, Ukraine}
}
\date{\today}
\maketitle
\begin{abstract}
\noindent
Inspired by previous extensive numerical studies of a cell fluid model with Curie-Weiss interactions, we concentrate on some analytically tractable special cases in its description.
The key ingredient of the model is a competition between global attraction and local repulsion interactions between particles with coupling constants $J_1$ and $J_2$, respectively.
We provide analytical results in several limiting cases, including the ideal-gas limit $J_1=J_2=0$ and the strong-repulsion limit $J_2\gg J_1$.
For $J_2\gg J_1$, a detailed analytical study is presented. We derive
explicit expressions for the critical point parameters, the equation of state, and
the binodal and spinodal curves in closed form. The equation of
state is found to be in full agreement with that of the van der Waals
lattice gas, and the order parameter satisfies the standard Curie-Weiss equation.
In a neighborhood of the critical point, a Landau expansion is shown to have the same form and symmetry as that of the classical lattice gas  within the mean-field approximation.
Moreover, based on the explicit knowledge of a few leading terms in the
asymptotic expansion of the deformed exponential function governing the
physics of the cell model, we extend its validity range to include the
marginal case of thermodynamic stability, $J_1=J_2$. In particular, this
extension makes a consideration of the ideal-gas limit $J_1=J_2=0$
formally legitimate. For the generic marginal case $J_1=J_2\ne0$
systematically avoided in previous works, we present numerical data and phase diagrams that augment their findings for $J_2>J_1$.

\end{abstract}


\section{Introduction}

The study of simple classical fluids and the phase transitions they exhibit is a cornerstone of statistical mechanics. The cell fluid model with Curie-Weiss interactions, introduced by Kozitsky et al. \cite{KKD20}, hereafter referred to as CFM, provides a tractable framework for modeling such systems.
The main characteristic features of CFM are the following.
The system's volume is divided into a macroscopically large number of non-overlapping cells of small finite size.
Any two particles in the system are influenced by a global mean-field-like attraction independently of their positions.
At short distances, however, they experience a repulsion when occupying the same cell, while any cell may be multiply occupied without restriction.
The thermodynamic stability of the system is ensured when the repulsion dominates over attraction, thus preventing the system's collapse.

The interplay between the attractive and repulsive two-particle interactions leads to a non-trivial phase behavior.
Extensive numerical investigations have demonstrated the existence of an infinite sequence of first-order transition lines, each terminating at a critical point \cite{KKD18,KD22,Retal25}.
This physical picture is reminiscent of the occurrence of multiple first-order phase transitions and associated critical points observed in several other systems, including a generalized Curie--Weiss spin model \cite{EE88}, cluster crystals \cite{NL11,WS13}, ultrasoft-particle systems \cite{MMC23}, and the one-dimensional generalized exponential
model of index~4 (GEM-4) \cite{Pre14,PGT15}.
Recently \cite{RDKPS26}, notable connections have been established between certain qualitative features of the CFM and multiple-occupancy quantum systems.

In CFM, the statistical occupation of a cell by $n$ particles is governed by  a discrete Gauss-Poisson probability distribution function \cite[(2.21)]{KKD20}, \cite{DSh24}
\begin{equation}\label{GPD}
p_{GP}(n ;z,r)=\left[K(r;z)\right]^{-1}\frac{{\rm e}^{zn}}{n!}\,
\mbox{e}^{-\frac 12\,rn^2}
\end{equation}
with support on $\mathbb N_0$ and parameters $z\in\mathbb R$ and $r\in\mathbb R_+$.
The normalization of this probability mass function is a special function given by the infinite sum
\begin{equation}\label{R}%\tag{A2}
K(r;z)=\sum_{n=0}^\infty\frac{{\rm e}^{zn}}{n!}\,{\rm e}^{-\frac 12\,rn^2}.
\end{equation}
It appears directly in the single-integral representation of the CFM's grand-canonical partition function \cite{KKD20,KD22,DSh24} (see also Section~\ref{RER}), and is thus a crucial component in its mathematical description and the derivation of the equation of state.

Apart from the sum $K(r;z)$ alone, several lowest-order moments of the distribution \eqref{GPD} are needed for actual calculations,
\be\label{BEM}
M_m(r;z)=\frac{K_m(r;z)}{K(r;z)}\mm{where}
K_m(r;z)=\sum_{n=0}^\infty\;n^m\;\frac{\e^{zn}}{n!}\e^{-\frac 12\,rn^2},\quad\quad m\in\mathbb N.
\ee
Of primary importance is the first moment $M_1(r;z)$,
\be\label{M1}
M_1(r;z)=\frac{K_1(r;z)}{K(r;z)}=\frac\partial{\partial z}\,\ln K(r;z)=\bn\,,
\ee
where $\bn$ denotes the mean value of the cell occupancy number $n$ with respect to the distribution function \eqref{GPD}.
%with appropriate physical parameters $r$ and $z$.
The sequence of moments $M_m(r;z)$ is also of mathematical interest beyond its physical applications.

The function $K(r;z)$ is structurally identical to the deformed exponential function appearing in \cite[(2.33)]{SS05} (see also \cite[(1.3)]{Sokal12}) in the context of the repulsive lattice gas and independent-set polynomials. This connection (overlooked in \cite{DSh24}) can be easily established through specific parameter redefinitions.

The deformed exponential function is known to have a periodic asymptotic behavior in the limit equivalent to $z\to\infty$, a phenomenon discovered by de Bruijn \cite{DeB49,DeB53}, with oscillations manifesting at order $O(1)$.
Graphical illustrations of the oscillating behavior of $K(r;z)$ and related functions can be found in \cite{DSh24}. Explicit asymptotic formulas in terms of the Jacobi theta function $\vartheta_3$ are derived in \cite{GS25}.
In fact, the presence of oscillations in $K(r;z)$ leads to the appearance of an infinite sequence of critical points in CFM, as discussed and graphically illustrated in \cite[Sec.~4]{Retal25}.

Although the well-known Laplace's method is directly applicable to the integral representation of the CFM's grand partition function in the thermodynamic limit,
there has been a notable lack of explicit analytical results for this model.
An exception is a quite general proof \cite{KKD20} of existence of a phase transition in CFM.
The main reason was the complexity of the function $\ln K(r;z)$ in the integrand, which complicates the location and analysis of extrema required by Laplace's method.

The current paper aims to fill this gap by presenting analytical calculations for several important physical scenarios, particularly in regimes that are analytically tractable or represent critical boundaries. We examine the model in well-understood physical limits to verify its consistency and explore the physics at the edge of thermodynamic stability.
%\bigskip

Our primary contributions include:

1. An extension of the validity range of the model's description to include the marginal case of thermodynamic stability, given by the equality of attraction and repulsion coupling constants, $J_1=J_2$.
This is achieved by using the first few terms of the asymptotic expansion of the function $\ln K(r;z)$ at large arguments $z$.
We show that the grand-canonical partition function of the CFM is mathematically convergent at the boundary $J_1=J_2$, which was systematically avoided in previous numerical treatments.
We support this theoretical result with new numerical data for the generic marginal case $J_1=J_2\ne0$ and phase diagrams in (density - pressure) and (density - temperature) planes.

2. Recovering the well-known classical result in the ideal-gas limit $J_1=J_2=0$, where no interactions are present in the system.
This serves as a necessary reference point for the model's consistency.

3. A detailed analytical treatment of the strong-repulsion limit
$J_2\gg J_1$. We derive explicit closed-form expressions for the
critical point parameters, the equation of state, and the binodal and spinodal
curves. Our analytical results are supported by graphical illustrations. The equation of state is shown to agree with that of the van
der Waals lattice gas \cite{PS15,FVbook}, and the order parameter
is found to satisfy the standard Curie-Weiss equation.
An exact symmetry of the binodal curves with respect to the reduced
critical density $\bar n_c=1/2$ is established in the present approximation.
A Landau expansion is shown to be consistent with that of the classical lattice gas in the mean-field approximation.

\section{The cell fluid model: basic definitions}
The object of this work is a cell model of a fluid with two-particle Curie-Weiss-type interactions introduced in \cite{KKD20,KK16}.
In these references, it was shown that multiple thermodynamic phases can exist in the system at sufficiently low temperatures.
Extended studies of this model have been subsequently done in \cite{KKD18,KD22}, and, more recently in \cite{DKPP26,Retal25,RDKPS26}. In this section, we introduce the CFM and related notation closely following \cite{KKD20}.

\subsection{Two-point interactions and their energy}\label{REX}

We consider an open classical many-particle system occupying a fixed macroscopic region of volume $V$ in three space dimensions, to be described within the grand canonical ensemble. The system's volume is partitioned into ${\mathcal N}$ non-overlapping congruent cubic cells $\Delta_\ell$\, ($\ell=1,...,{\mathcal N}$) of volume $v=V/{\mathcal N}$. The system consists of a variable number $N\in\mathbb N_0$ of point-like particles with three-dimensional coordinates $\{x_1,...,x_N\}$. The particles can occupy any of the cells $\Delta_\ell$ in the volume $V$, with multiple occupancy permitted.

The two-point interaction potential of particles with coordinates $x_i$ and $x_j$ is given by
\be\label{CV}
\Phi_{\mathcal N}(x_i,x_j)=-\,\frac{J_1}{{\mathcal N}}+J_2\sum_{\ell=1}^{\mathcal N} I_\ell(x_i)\,I_\ell(x_j)
\ee
where $J_1$ and $J_2$ are coupling constants corresponding to distinct physical mechanisms. In \eqref{CV},
\vspace{-1mm}
\bi\itemsep0.1mm
\item[$\bm\cdot$] $J_1>0$ measures the strength of a mean-field-like global uniform attraction between any two particles in the volume $V$, independent of their specific positions. The interaction is scaled by $1/{\mathcal N}$ to ensure the existence of a thermodynamic limit. This scaling, providing the energy growth $\sim\mathcal N$ when $\mathcal N\to\infty$, is known as the Kac normalization, see e.g. \cite[Sec.~2.1]{Campa}, \cite{Kastner25} and references therein;
\item[$\bm\cdot$] $J_2>0$ controls the local repulsion strength between the particles inside any cell $\Delta_\ell$. The corresponding term in \eqref{CV} may be seen as modeling the physically reasonable soft-core inter-particle repulsions~\cite{Likos2001};
\item[$\bm\cdot$] $I_\ell(x_i)$ is the indicator function for the cell $\Delta_\ell$ defined as
\vspace{-2mm}
\bc
$I_\ell(x_i)=1$\quad if\quad $x_i\in\Delta_\ell$,\qquad
$I_\ell(x_i)=0$\quad if\quad $x_i\notin\Delta_\ell$\,.
\ec
\ei

The total energy of the system containing $N$ particles, $W_{\mathcal N}^{(N)}$, is given by a sum of pair interaction energies given in \eqref{CV}.
Following \cite{KKD20}, we include the self-interaction terms $(i=j)$ in $W_{\mathcal N}^{(N)}$ accepting that this does not affect the physics of the model in the thermodynamic limit. The total energy is thus
\be\label{OV}
W_{\mathcal N}^{(N)}=\frac12\sum_{i,j=1}^N\Phi_{\mathcal N}(x_i,x_j)=
-\frac12\,\frac{J_1}{\mathcal N}\,N^2+\frac12\,J_2\sum_{\ell=1}^{\mathcal N}\,n_\ell^2\,,
\ee
where $n_\ell$ is the number of particles occupying a given cell $\Delta_\ell$\,:
\be\label{ORC}
n_\ell=\sum_{i=1}^N I_\ell(x_i)\,.
\ee

The presence of two different terms with coupling constants $J_1$ and $J_2$ in \eqref{OV} gives rise to a competition between the global Curie-Weiss attraction and the  local, short-range repulsion. This competition leads to a frustration in the system, responsible for the non-trivial phase behavior discussed in \cite{KD22}.

According to a rigorous result by Ruelle \cite{Ruelle70} (as quoted in \cite{KKD20}), the thermodynamic stability of the system is provided by the condition
\be\label{RUE}
\int_Vdx\,\Phi_{\mathcal N}(x,\tilde x)>0\qquad\forall\,\tilde x\,\in V\,,
\ee
where the integration runs over the three-dimensional volume $V$ of the system, and we use a short-hand notation $dx=d^3x$. For the potential $\Phi_{\mathcal N}(x_i,x_j)$ in \eqref{CV}, the condition \eqref{RUE} implies the requirement
$J_2>J_1$ \cite[(2.3) -- (2.4)]{KKD20}, see also \eqref{f} below.
\label{ppr}
This inequality ensures that short-range repulsive interactions dominate over long-range attraction at small scales, thus preventing the system collapse.

\section{Thermodynamics}\label{RET}

In this section, we introduce the thermodynamic framework for our study of the CFM.
Its subsequent analysis will be carried out within the grand canonical ensemble.
We present the most general thermodynamic relations and their implications for the specific model under consideration.

\subsection{General relations}\label{RET1}

All thermodynamic properties of the system derive from the grand-canonical partition function (GPF), see e.g. \cite[(2.4.6)]{HansenMcDonald13}, at the absolute temperature $T_{\rm phys}$,
\be\label{ZGR}
\Xi_{\mathcal N}=\sum_{N=0}^\infty\,\frac{\zeta^N}{N!}\;Z_{\mathcal N}^{(N)}
\ee
involving the configuration integral
\be\label{zhh}
Z_{\mathcal N}^{(N)}=\int(dx)^N\exp\big[{-\beta W_{\mathcal N}^{(N)}}\big]\,.
\ee
The inverse temperature $\beta$ and activity (fugacity) $\zeta$ are given by
\be\label{LA3}
\beta=\frac1{k_B\,T_{\rm phys}}\MM{and} \zeta=\frac{\e^{\beta\mu_{\rm phys}}}{\Lambda^3}\,.
\ee
Here, $k_B$ is the Boltzmann constant, $\mu_{\rm phys}$ is the physical chemical potential, and $\Lambda$ is the de~Broglie thermal wavelength
\be\label{Lam}
\Lambda=\hbar\,\sqrt{\frac{2\pi\beta}m}
\ee
with $\hbar$ being the reduced Planck constant and $m$ the particle mass.
The two independent fundamental thermodynamic variables in the problem are the inverse temperature
$\beta\ge0$ and the physical chemical potential $\mu_{\rm phys}\in\mathbb R$.
\vskip 1mm

The mean values of physical quantities are defined as statistical averages over the grand canonical ensemble via
\begin{equation}
\overline{(\,\cdots)}\,=\frac{1}{\Xi_{\mathcal N}}\,\sum_{N=0}^{\infty}(\,\cdots)
\frac{\zeta^N}{N!}\,Z_{\mathcal N}^{(N)}.
\end{equation}
In particular, the observable mean particle number in the system is given by
\begin{equation}\label{ndef}
\overline N=\,\frac{1}{\Xi_{\mathcal N}}\,\sum_{N=0}^{\infty}\,N\,\frac{\zeta^N}{N!}\,Z_{\mathcal N}^{(N)}\,=
\beta^{-1}\,\frac{\partial\ln\Xi_{\mathcal N}}{\partial\mu_{\rm phys}}\,.
\end{equation}
Furthermore, it is useful to introduce the mean particle number per cell $\bn_\ell=\bn$, or the statistical cell occupancy stemming from the initial definition \eqref{ORC}, identical for all cells. In terms of the overall mean particle number $\overline N$, it is simply given by
\begin{equation}\label{nc}
\bn=\frac{\overline N}{\mathcal N}\,.
\end{equation}
In other words, the mean cell occupancy $\bn$ is equal to the reduced particle density, denoted by $\rho^*$ in \cite[Sec. 3.1]{Retal25}.
In Section \ref{RER}, we shall establish the consistency of the thermodynamic definition \eqref{nc} with its probabilistic counterpart \eqref{M1}.
\vskip 1mm

\label{tdl}
In the currently studied cell model, the thermodynamic limit corresponds to the limit
${\mathcal N}\to\infty$ of the total cell number ${\mathcal N}$ in the above definition of the GPF and the related mean values of thermodynamic quantities.
At a fixed cell volume $v$, the limit ${\mathcal N}\to\infty$ drives the system's volume $V={\mathcal N}v$ to infinity.
In the thermodynamic limit, the grand potential $\Omega$ is given by
\be
\Omega=-k_{\rm B}T_{\rm phys}\ln\Xi\,,\MM{where}\ln\Xi={\mathcal N}\lim_{{\mathcal N}\to\infty}{\mathcal N}^{-1}\,\ln\Xi_{\mathcal N}\,
\ee
and $\Xi_{\mathcal N}$ is defined in \eqref{ZGR}.

The pressure $P$ is defined by the well-known general thermodynamic formula
\begin{equation}\label{eos}
PV=-\Omega=k_{\rm B}T_{\rm phys}\ln\Xi\,.
\end{equation}
Dividing the last equation by the number of cells ${\mathcal N}$, we obtain more specifically for CFM,
\begin{equation}\label{eod}
\beta Pv=\lim_{{\mathcal N}\to\infty}{\mathcal N}^{-1}\,\ln\Xi_{\mathcal N}\,.
\end{equation}
In the present framework of the grand canonical ensemble, this relation is actually the equation of state for the system under consideration.

\subsection{Dimensionless quantities}\label{hdq}

In this section, we introduce our notations and definitions for  dimensionless physical quantities to be used throughout the paper.

In the combination $\beta W_{\mathcal N}^{(N)}$ appearing in \eqref{zhh} with $W_{\mathcal N}^{(N)}$ given by \eqref{OV}, we encounter two natural dimensionless parameters, $p\equiv\beta J_1$ and $r\equiv\beta J_2$.
Following \cite{KKD20}, we choose the quantity $p$ to represent the  dimensionless inverse temperature.
The parameters $p$ and $r$ are proportional to one another, and, basically, each of them could be chosen to be a temperature-like variable.
Their ratio is
\be\label{f}
\frac r p=\frac{J_2}{J_1}\equiv f,
\ee
and the Ruelle's stability condition \eqref{RUE} requires that $f>1$.

The dimensionless combination $\beta\mu_{\rm phys}$ appearing in the fugacity \eqref{LA3} is then represented as
$\beta\mu_{\rm phys}=\beta J_1{\cdot}\,\mu_{\rm phys}/J_1\equiv p\,\mu_0$,
where we define the dimensionless  chemical potential $\mu_0\in\mathbb R$ via $\mu_0\equiv\mu_{\rm phys}/J_1$.

Next, we extract the temperature dependence from the de~Broglie wavelength \eqref{Lam} via
\be\label{lam}
\Lambda=\sqrt p\,\lambda\,,\MM{where}\lambda\equiv\hbar\,\sqrt{\frac{2\pi}{mJ_1}}
\ee
is a temperature-independent constant with dimension of length. This allows us to define a dimensionless volume of a cell as
\be\label{vest}
v^*=\frac v{\lambda^3}\,.
\ee

Finally, for the dimensionless combination $\beta Pv$ appearing in the equation of state
\eqref{eod} we write $\beta Pv=pP^*$ where the
dimensionless  pressure appears as in \cite[Sec. 3.1]{Retal25}:
\be\label{hdp}
P^*\equiv P\,\frac v{J_1}\,.
\ee
The equation of state itself is given thus in the form (cf. \eqref{eod})
\be\label{eons}
pP^*=\lim_{{\mathcal N}\to\infty}{\mathcal N}^{-1}\,\ln\Xi_{\mathcal N}\,.
\ee
The right-hand side of this equation will be made explicit in the next section.

In presenting the CFM's phase diagrams in Sections \ref{PDf1} and \ref{ph}, we shall also use the dimensionless temperature variable
\be\label{TTD}
\T=p^{-1}=k_B T_{\rm phys}/J_1\,.
\ee

\section{Integral representations of the grand-canonical partition
function}\label{RER}

An explicit calculation detailed in \cite{KKD20} yields a single-integral representation
for the CFM's grand-canonical partition function $\Xi_{\mathcal N}=\Xi_{\mathcal N}(p,r,\mu_0,v)$.
Adopting a parametrization slightly differing from that of \cite{KKD20},
we reproduce it in the form
\be\label{Y}
\Xi_{\mathcal N}=\sqrt{\frac {\mathcal N}{2\pi p}}\,\int_{-\infty}^\infty dy\,
\exp\Big\{{\mathcal N} \Big[-\frac{y^2}{2p}\,+\,
\ln\sum_{n\ge0}\frac{v^n}{n!}\e^{(y+p\mu_0)n-\frac 12\,r n^2}\Big]\Big\}\,.
\ee
The infinite sum in the integrand represents the function $K(r;z)$
introduced in \eqref{R} with a shifted argument:
\be\label{KK}
\sum_{n\ge0}\frac{v^n}{n!}\e^{(y+p\,\mu_0)n-\frac 12\,r n^2}=K(r;z)\,,
\MM{where}z=y+p\,\mu_0+\ln v\,.
\ee

The thermodynamic limit (see p.~\pageref{tdl}) is given by
${\mathcal N}\to\infty$, and the evaluation of the integral over $y$
in \eqref{Y} proceeds using the standard Laplace's method
\cite{Fedoryuk89,Wong,FS}. This yields the asymptotically exact result
\be\label{KEY}
\lim_{{\mathcal N}\to\infty}{\mathcal N}^{-1}\ln\Xi_{\mathcal N}=
E(\bar y_{\rm max})\,,
\ee
where $E(\bar y_{\rm max})$ is the global-maximum value of the
function $E(y)=E(p,r,\mu_0,v;y)$ appearing in the exponent of \eqref{Y}:
\be\label{EE}
E(y)=-\frac{y^2}{2p}\,+\,
\ln\sum_{n\ge0}\frac{v^n}{n!}\e^{(y+p\mu_0)n-\frac 12\,r n^2}\,.
\ee
The position $\bar y_{\rm max}$ of the global maximum is defined as
the solution to the extremum condition $E'(y)=0$, subject to the stability requirement
$E''(\bar y_{\rm max})<0$ (which excludes all minima and inflection
points), and the condition $E(\bar y_{\rm max})>E(y)$ for all $y\ne\bym$.
Since the condition $E'(y)=0$ plays a central role in the following,
we write it explicitly using \eqref{EE} and \eqref{M1}:
\be\label{EXE}
E'(y)=-\frac yp+M_1(r;z)=0\,.
\ee

The application of Laplace's method provides, via \eqref{KEY}, the explicit form for the right-hand side of the equation of state \eqref{eons}. Thus, the dimensionless pressure $P^*$, which depends on the reduced temperature through $p=1/\T$ and on the chemical potential $\mu_0$, is given by the asymptotically exact equation of state
\be\label{ERR}
pP^*=E(\bar y_{\rm max})\,.
\ee
A more detailed exposition of the formalism is available in \cite[Sec.~3]{Retal25}.

Combining the thermodynamic definition of the mean reduced particle density $\bn$ given by \eqref{ndef} -- \eqref{nc} with \eqref{KEY} yields
\be\label{LAL}
\bn=\beta^{-1}\,\frac\partial{\partial\mu_{\rm phys}}\, \lim_{{\mathcal N}\to\infty}{\mathcal N}^{-1}\,\ln\Xi_{\mathcal N}=
p^{-1}\,\frac\partial{\partial\mu_0}\,E(\bar y_{\rm max})\,.
\ee
In performing the differentiation with respect to $\mu_0$ in \eqref{LAL}, we have to account for the implicit dependence of $\bym$ on $\mu_0$ which is defined by \eqref{EXE}. Via the chain rule we have
\be
\bn=p^{-1}\,\frac{\partial E(\bar y_{\rm max})}{\partial\mu_0}+
p^{-1}\,\frac{\partial E(\bar y_{\rm max})}{\partial\bym}\,
\frac{\partial\bym}{\partial\mu_0}\,.
\ee
The second term vanishes identically due to the extremum condition $E'(\bym)=0$, and we remain with (cf. \cite[(2.26)]{KKD20}, \cite[Sec. 3.5]{Retal25})
\be\label{ncy}
\bn=M_1(r;z(\bym))=\frac\bym p\,,
\ee
where the second equality follows from \eqref{EXE}.
The equation \eqref{ncy} agrees with the probabilistic definition \eqref{M1}.

Finally, we note that in \eqref{KK}, the term $\ln v$
merely shifts the quantity $p\mu_0$ (denoted as $\mu_0$ in \cite[p.~5]{KKD20}). This structure indicates that the specific value of the  cell volume $v$ cannot affect the physics of the system, a conclusion supported by the extensive study of the CFM's phase behavior in \cite{DKPP26}.
This indicates that the explicit dependence on $v$ can be eliminated, as shown in Section \ref{REM} through a redefinition of the chemical potential.

By contrast, the interplay between attraction and repulsion governed by the parameter $f=r/p$ is interesting \cite{Petal25}, and its role will be analyzed in the following sections.

\subsection{An alternative form of the grand-canonical partition function}
\label{REA}

The form of the argument of the function $K$ in \eqref{Y} -- \eqref{KK} suggests a shift of the integration variable in \eqref{Y} via $y=z-p\mu_0-\ln v$ first performed in \cite{KD22}.
Thus, the main integral becomes
\be\label{Z}
\Xi_{\mathcal N}=\sqrt{\frac {\mathcal N}{2\pi p}}\int_{-\infty}^\infty\!dz
\exp\Big\{{\mathcal N} \Big[-\frac1{2p}(z-p\,\mu_0-\ln v)^2+\ln K(r;z)\Big]\Big\},
\ee
where the function $K(r;z)$ appears as defined in \eqref{R}.
In comparison to \eqref{Y}, here the function $\ln K(r;z)$ contains \emph{only one}
physical parameter, $r=\beta J_2$, where $J_2$ refers to the local repulsion energy.
All remaining "physics" enters the first simple quadratic term.

The evaluation of the integral \eqref{Z} via Laplace's method proceeds along the lines very similar to that outlined in the preceding section, and physical consequences are discussed in \cite{KD22}.
\vskip1mm

However, at this point we would like to escape the usual routes and address a novel aspect of the theory: what is the behavior of the integral \eqref{Z} at the edge of the Ruelle's thermodynamic stability condition \eqref{RUE}, that is when the attraction and repulsion coupling constants are equal?
In other words, if we set $J_1=J_2$ (or $r=p$), would it lead to a divergence in \eqref{Z}? Or would the integral remain well-defined as in the general case $J_1<J_2$?

\section{The issue of convergence at the boundary of thermodynamic  stability} \label{SIS}

The thermodynamic stability condition $f=J_2/J_1>1$ (or $r>p$ at any finite temperature) is crucial for ensuring that any integral representation for the GPF $\Xi_{\mathcal N}$ converges. The marginal case $J_1=J_2$ requires a special consideration.

We start by completing the square in the exponents of summands in $K(r;z)$ (see \eqref{R}), by writing
\be\label{CS}
zn-\frac 12\,rn^2=\frac{z^2}{2r}-\frac r2\,\Big(n-\frac zr\Big)^2.
\ee
Here, the first term on the right-hand side is the maximal value of the quadratic form on the left, for any $n\ge0$. Hence, for the GPF $\Xi_{\mathcal N}$ in \eqref{Z} we can write
\be\label{YR}
\Xi_{\mathcal N}\propto\e^{-\frac {\mathcal N}{2p}\,(p\,\mu_0+\ln v)^2}\!\int_{-\infty}^\infty dz
\exp\Big\{{\mathcal N}\Big[-\frac{z^2}{2p}\,\left(1-b\right)
+(\mu_0+p^{-1}\ln v)\,z+\ln H(r;z)\Big]\Big\},
\ee
where we define the ratio $\displaystyle{b:=\frac pr=\frac{J_1}{J_2}=f^{-1}}$
in the range $b\in[0,1]$ (cf. \eqref{f}) and the modified $K$-sum (cf. \eqref{R}) with moderated asymptotic growth,
\be\label{HR}
H(r;z):=\sum_{n\ge0}\,\frac1{n!}\e^{-\frac r2\,\left(n-\frac zr\right)^2}=
\e^{-z^2/(2r)}K(r;z)\,.
\ee

While the function $\ln K(r;z)$ roughly behaves as
\be\label{o}
\ln K(r;z)\sim\frac{z^2}{2r}+o(z^2) \MM{when} z\to\infty,
\ee
this implies that $\ln H(r;z)\sim o(z^2)$ in the same limit.
For the integral in \eqref{YR}, this means that its convergence at large $z$
is controlled by the first quadratic term $\propto-z^2(1-b)$, vanishing when $b=1$.
In particular,
\vspace{-2mm}
\bi\itemsep-1mm
\item[$\bm\cdot$]
the inequality $b<1$ ensures the convergence of the integral over $z$ in \eqref{YR},
in agreement with \eqref{RUE} and \eqref{f};
\item[$\bm\cdot$]  with $b>1$, the integral would diverge, which corresponds to an unphysical situation resulting in a collapse of the system;
\item[$\bm\cdot$] the marginal case $b=1$ requires a special consideration taking into account the leading asymptotic behavior of the function $\ln H(r;z)$ at large $z$.
\ei
In fact, in the absence of the $\propto\,z^2$ term in \eqref{YR}, the
convergence of the integral at large $z$ is controlled by the leading
term of the asymptotics of $\ln H(r;z)$, that is, by the sub-leading $o(z^2)$
term in the asymptotic expansion of $\ln K(r;z)$ implied by \eqref{o}.
The knowledge of this term will allow us to answer the question about the convergence
of the integral representation \eqref{Z} for the GPF $\Xi_{\mathcal N}(p,r,\mu_0,v)$ in the delicate marginal case $r=p$.
To determine the needed term(s) of the asymptotic expansion of $K(r;z)$ as $z\to\infty$, one has to resort to elaborate techniques involving the discrete Laplace's method described in \cite{deBruijn,BO78,FS,Paris11}.

\subsection{Some elementary properties of the function $K(r;z)$ and its asymptotic behavior at large $z$}\label{RRR}

The function $K(r;z)$ is well-defined for any $r\ge0$ and $z\in\mathbb R$. At $r=0$, the exact result of summation in \eqref{R} is $K(0;z)=\exp(\e^z)$, while in the limit $r\to\infty$ only the first term of the sum survives and thus we have $K(\infty;z)=1$.
The same happens for any finite $r$ and $z\to-\infty$: in this limit, $K(r;z\to-\infty)=1$.
Thus, for any $0\le r<\infty$, the function $K(r;z)$ lies within the bounds
$1\le K(r;z)\le\exp(\e^z)$ for all $z\in\mathbb R$, and therefore $0\le\ln K(r;z)\le\e^z$.
This bound is weaker than $K(r;z)\le\e^{z^2/(2r)+1}$, used in \cite{KKD20} for $r>0$, but has the advantage of being valid uniformly for all $r\ge0$, including the $r=0$ case.

In our following consideration, the asymptotic behavior of $K(r;z)$ for $0<r<\infty$ and $z\gg1$ will play an important role, and we shall need its more detailed description.

In \cite[(21)]{DSh24}, the monotonic, non-oscillatory part of the asymptotics of the function $r\ln K(r;z)$ has been identified. Writing $K^{(\rm as)}(r;z)$ for the corresponding asymptotic approximation, we have
\be\label{ARS}
r\ln K^{(\rm as)}(r>0;z\gg1)\sim
\frac12\,z^2-z(\ln\xi-1)+\frac12\,\ln^2\xi-\frac r2\,\ln z+
O\Big(\frac{\ln\xi}z\Big)\,,
\ee
where we use the notation $\xi\equiv z/r$.
The complete asymptotic formula involving the oscillating component, expressed in terms of the Jacobi theta function $\vartheta_3$ and overlooked in \cite{DSh24}, will appear in \cite{GS25}. For our present purposes, the knowledge of a few leading monotonic terms from \eqref{ARS} is sufficient.

\subsection{Convergence of the GPF's integral representation at the stability edge}\label{SII}

Now, possessing the required information on the large-$z$ behavior of the function
$\ln K(r;z)$, we are in a position to solve the convergence problem of the GPF's integral representation \eqref{Z} at $J_1=J_2$. Let us write its transformed version from \eqref{YR} in the form
\be\label{BI}
\int_{-\infty}^\infty\!dz\e^{{\mathcal N}\,E(p,r;z)}\MM{where}
E(p,r;z):=-\frac{z^2}{2p}\,\left(1-b\right)+(\mu_0+p^{-1}\ln v)\,z+\ln H(r;z)\,,
\ee
and the function $H(r;z)$ is defined in \eqref{HR}.
We set here $r=p$, which implies $b=1$. The quadratic term in $E(p,r;z)$ disappears, and the integrand now involves the function
\be\label{E0}
E(p,p;z)=(\mu_0+p^{-1}\ln v)\,z+\ln H(p;z)\,.
\ee

The definition of $H(r;z)$ in \eqref{HR} implies that the asymptotic behavior of $\ln H(p;z)$ at large $z$ is given by
\be
\ln H^{(\rm as)}(p;z\gg1)=\ln K^{(\rm as)}(p;z\gg1)-\frac{z^2}{2p}\sim
-\frac zp\,(\ln z-\ln p-1)+O(\ln^2z)\,,
\ee
where the expression on the right-hand side results from the truncated asymptotic
estimate \eqref{ARS} with the replacement $r\mapsto p$. Thus, for large enough $z$ we have
\be\label{E01}
E(p,p;z\gg1)\sim-\,\frac zp\,\ln z+
\,\frac zp\left(p\,\mu_0+\ln v+\ln p+1\right)+O(\ln^2z)\,.
\ee
We see that the negative next-to-leading asymptotic term $\sim-z\ln z$ of the function
$p\ln K^{(\rm as)}(p;z \gg1)$ from \eqref{ARS} dominates over the linear $O(z)$-term from \eqref{E0} and guarantees the convergence of the integral \eqref{BI} at $z\to+\infty$.

This means that at $J_1=J_2$, the grand partition function of CFM is given by a convergent integral and is thus expected to yield meaningful physical results. Its numerical evaluation can be safely performed to include this marginal case, thereby extending the results of earlier works \cite{KKD20,KKD18,KD22} where only values of $b=f^{-1}$ strictly smaller than unity were considered. Examples of the latter can be found in \cite[Table 1]{KKD20}, \cite[p. 249]{KKD18}, and more recently in \cite[Table 2]{DKPP26} and \cite[Table 2]{Petal25}.

Before presenting the new numerical data and phase diagrams corresponding to the interacting system with $J_1=J_2\ne0$, we briefly address its trivial limiting case $J_1=J_2=0$.

\section{The ideal gas limit}\label{SSS}

The existence of the GPF established in Section~\ref{SII} for the boundary of thermodynamic stability $J_1=J_2$ makes formally legitimate the consideration of the further limiting case, $J_1=J_2=0$.
It corresponds to the ideal gas, where particles do not interact, and requires taking the limits $p\to0$ and $r\to0$ in the GPF's integral representations \eqref{Y} or \eqref{Z} while keeping the inverse physical temperature $\beta$ finite.

We start with equation \eqref{Y} and remove the problematic factor $1/\sqrt p$ in front of the integral by changing the integration variable via $y/\sqrt p=t$. Taking into account the presence of de~Broglie thermal wavelength $\Lambda$ in the activity \eqref{LA3}, we obtain
\be\label{T}
\Xi_{\mathcal N}=\sqrt{\frac {\mathcal N}{2\pi}}\int_{-\infty}^\infty\!dt
\exp\Big\{{\mathcal N} \Big[-\frac{t^2}2+\ln\sum_{n\ge0}\frac{(v/\Lambda^3)^n}{n!}\,
\e^{(t\sqrt p\,+\beta\mu_{\rm phys})n-\frac 12\,r n^2}\Big]\Big\}.
\ee
The limits $p\to0$ and $r\to0$ can be safely taken, leaving a simple exponential sum in the integrand of \eqref{T}. The resulting integral becomes Gaussian, yielding the ideal-gas GPF
\be\label{TRE}
\Xi_{\mathcal N}^{(id)}(\beta,\mu_{\rm phys})=
%\sqrt{\frac {\mathcal N}{2\pi}}\int_{-\infty}^\infty\!dt\exp\Big\{{\mathcal N}\Big[-\frac{t^2}2+\ln\sum_{n\ge0}\frac{(v/\Lambda^3)^n}{n!}\,\e^{\beta\mu_{\rm phys} n}\Big]\Big\}
\exp\left({\mathcal N}v\e^{\beta\mu_{\rm phys}}/\Lambda^3\right).
\ee
With ${\mathcal N}v=V$, we obtain
\be\label{IGL}
\ln\Xi_{\mathcal N}^{(id)}(\beta,\mu_{\rm phys})=\e^{\beta\mu_{\rm phys}}V/\Lambda^3\,,
\ee
in full agreement with the well-known classical result of \cite[p.~395, (A2.11)]{Hill56}.

Though uncomplicated,  this consistency check is important: it validates both the formalism used to analyse the CFM, and the formulation of the model itself.

In the next section, we report new numerical data and phase diagrams for the interacting CFM with equal coupling constants $J_1$ and $J_2$.

\begin{figure}[!]%htbp
	\centering
	\includegraphics[width=0.82\textwidth]{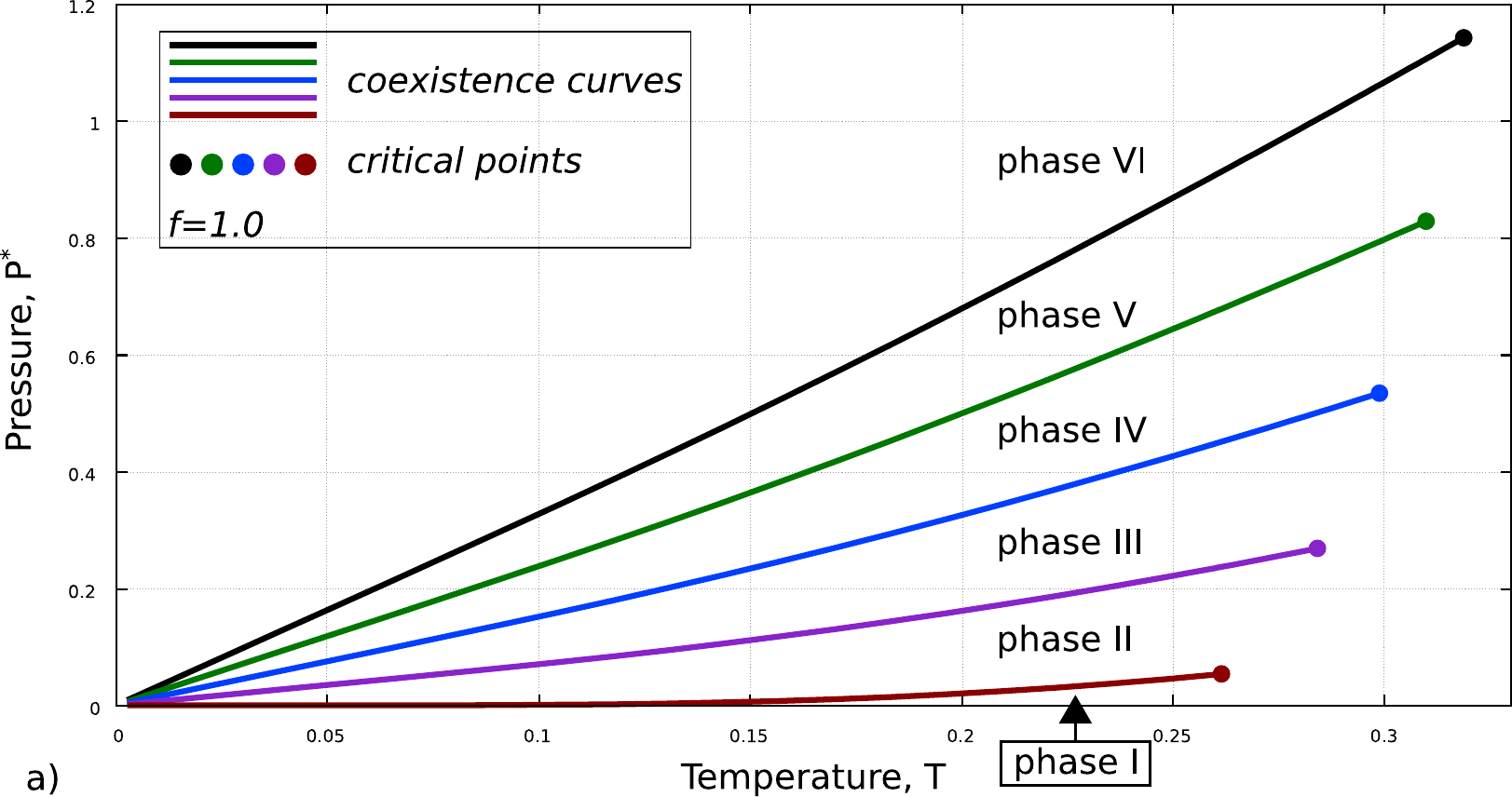}\\[4mm]
	\includegraphics[width=0.82\textwidth]{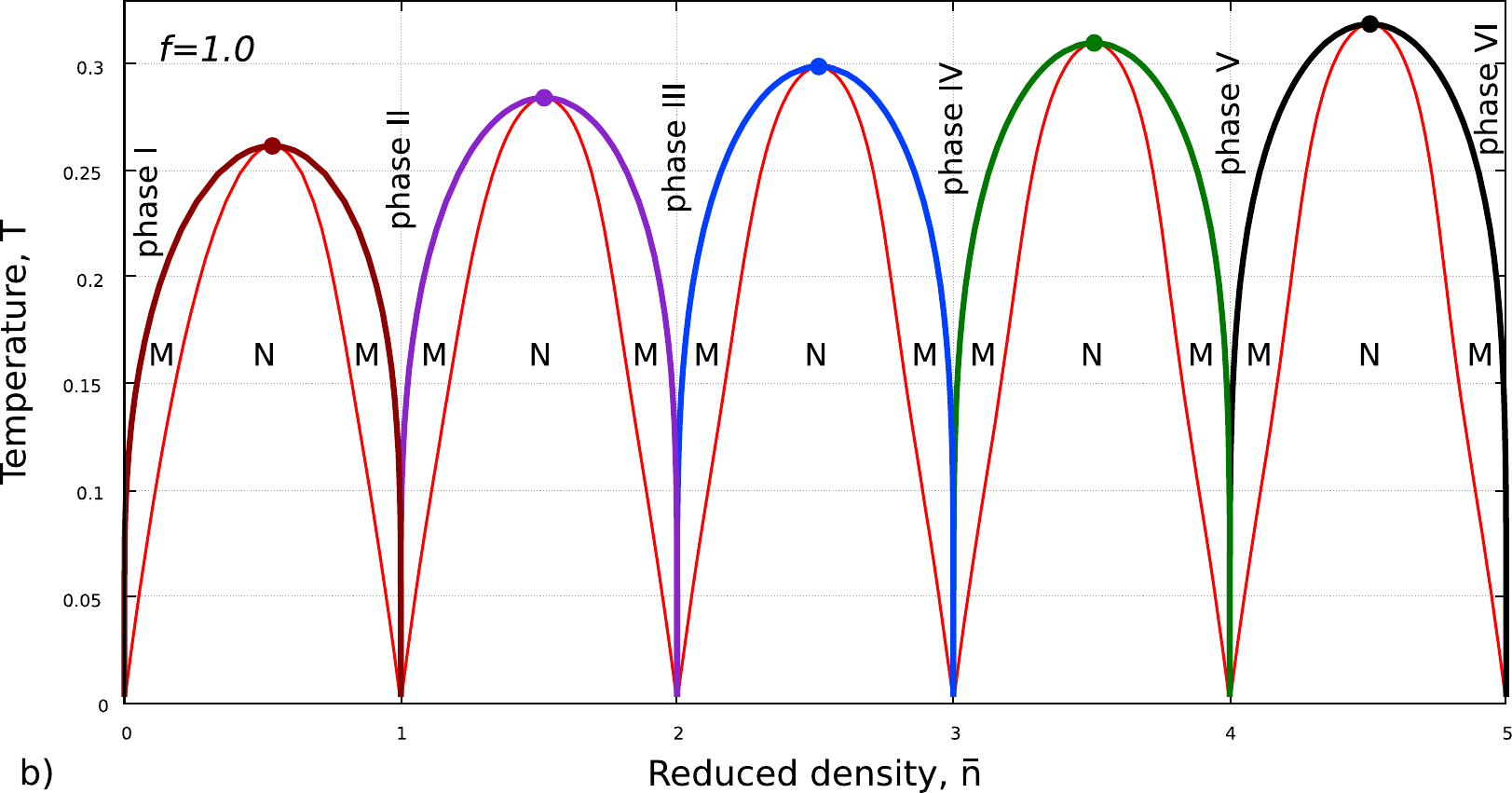}
\caption{Phase diagrams of the cell fluid model with Curie-Weiss-type
    interaction potential~\eqref{CV} with equal attraction and repulsion parameters, $J_1=J_2$.
    %for the repulsion-to-attraction ratio $f=1$.
\textit{Fig.~\ref{f1}a:} The $(P^*,\T)$ phase diagram showing the first five first-order phase transition lines separating six distinct phases with growing density. Each coexistence line terminates at its critical point (solid circle) on the right-hand side, labeled $k=1,\ldots,5$ in Table~\ref{tab1}.
\textit{Fig.~\ref{f1}b:} The $(\T,\bar{n})$ phase diagram displaying the phase coexistence lines (binodals) and boundaries of thermodynamically unstable regions (spinodals). The brown, purple, blue,
    green, and black solid curves indicate coexistence between
    phases~I--II, II--III, III--IV, IV--V, and V--VI, respectively.
    Spinodals are represented by thin red curves. Regions labeled by M and N denote metastable and unstable states. Critical point parameters are listed in Table \ref{tab1}.}
\label{f1}
\end{figure}
\begin{figure}[htbp]
	\centering
	\includegraphics[width=0.82\textwidth]{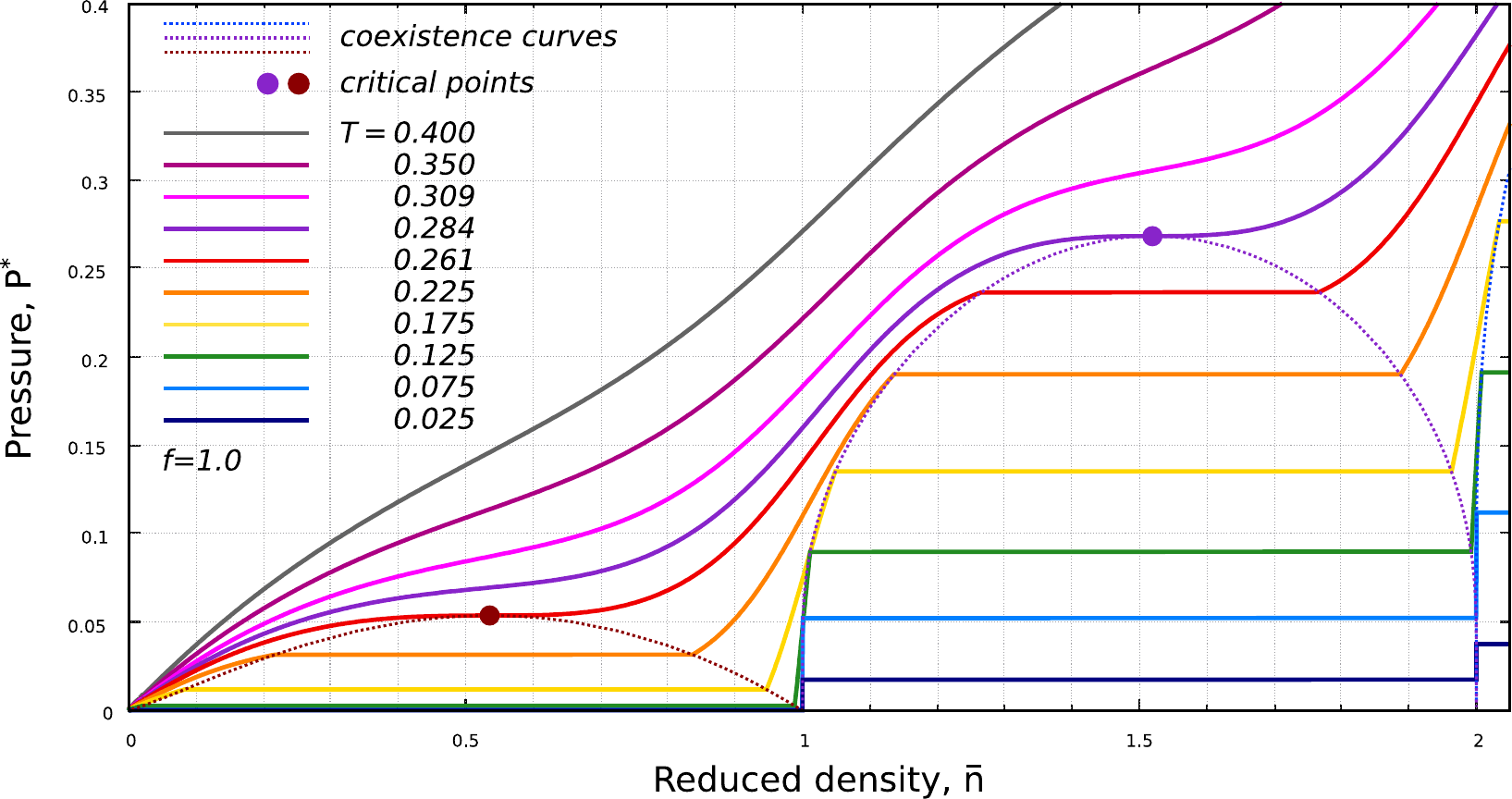}
	\caption{The $(P^*,\bar n)$ phase diagram of the cell fluid model with equal attraction and repulsion parameters, $J_1=J_2$, showing pressure isotherms (solid lines) at ten values of the reduced temperature $\T$ listed in the legend.
The first two coexistence lines (binodals) related to the three lowest-density phases are indicated by dotted curves. Their apex points correspond to the critical points labelled by $k=1$ and $k=2$ in Table~\ref{tab1}. The horizontal segments of the isotherms in the subcritical region indicate discontinuous density jumps accompanying the first-order phase transitions.}
	\label{f2}
\end{figure}

\section{Phase diagrams in the marginal case $J_1=J_2$}\label{PDf1}

Numerical analysis of the equation of state \eqref{ERR}, carried out along the usual lines of \cite{KKD20,KKD18,KD22} with $f=1$, yields the phase diagrams shown in Figures~\ref{f1}a), \ref{f1}b), and \ref{f2}. These three complementary representations provide a full picture of the phase behavior of the multiple-occupancy cell fluid model defined in \eqref{CV}--\eqref{OV} for the case of equal repulsion and attraction coupling constants, $J_1=J_2$. Together, these phase diagrams reveal a sequence of first-order phase transition lines separating distinct phases, displaying the associated coexistence and metastability regions, and the corresponding critical points. Although only the first five phase-transition and coexistence lines are shown in Fig.~\ref{f1} for clarity, and in Fig.~\ref{f2} only the first two, the complete theory allows, in principle, for an infinite sequence of such transitions and critical points associated with progressively increasing average cell occupancy number $\bar n$ \cite{KD22,Retal25}.

In the $(P^*,T)$ phase diagram shown in Fig.~\ref{f1}a, the five coexistence lines separate phases labelled by I,\ldots,VI. Each coexistence line terminates at a critical point located at its high-temperature end, the corresponding critical-point parameters are listed in Table~\ref{tab1}. The coexistence lines are ordered by pressure, exhibiting a monotonic increase in both critical temperature and critical pressure with increasing phase index. This indicates that phases with higher average occupancy numbers are thermodynamically stable at progressively higher pressures and remain stable in certain temperature ranges. The absence of intersections between coexistence lines implies that no triple points occur in the present theory \cite{DKPP26} (for a modification of the CFM that allows for the appearance of a triple point, see \cite{triple}).

Figure~\ref{f1}b presents the corresponding phase diagram in the $(\bar n,T)$ plane. The coexistence regions appear as a sequence of closed binodal domes centered approximately around half-integer values of the reduced density $\bar n$. As in Fig.~\ref{f1}a, the brown, purple, blue, green, and black curves correspond to coexistence between phases I–II, II–III, III–IV, IV–V, and V–VI, respectively. Each binodal curve reaches a critical point at its maximum, where the density difference between coexisting phases vanishes continuously. Inside every binodal dome, thin red curves represent the spinodal boundaries separating metastable and unstable states. The regions labeled by M correspond to metastable states, while those labeled by N enclose thermodynamically unstable states. Figure~\ref{f1} presents the marginal case, at $f=1$, of the series of analogous phase diagrams in \cite[Fig.~2]{DKPP26}.

Figure \ref{f2} displays the $(P^*, \bar n)$ phase diagram, illustrating the equation of state by ten selected isotherms. Solid curves represent the pressure isotherms, while the dotted curves indicate the binodal boundaries with respective critical points at their maxima. The color scheme is consistent with that of coexistence curves in Fig.~\ref{f1}. At temperatures below the respective critical values, the isotherms exhibit the non-monotonic behavior characteristic of first-order phase transitions (the van der Waals loops; cf. Fig.~\ref{f5}). These portions are replaced by horizontal plateaus representing the phase coexistence and density jumps between the coexisting phases. The width of these horizontal segments correspond exactly to the width of the first two coexistence domes shown in Fig.~\ref{f1}b. As the temperature rises toward $T_c^{(1,2)}$, the plateaus shrink continuously and disappear at the corresponding critical points. At high enough temperatures, the isotherms become smooth and monotonic. In both Figures~\ref{f1} and \ref{f2}, the dimensionless temperature $\T=1/p$ (see \eqref{TTD}) ranges from $0.01T_c^{(1)}\simeq0.0026$ up to the supercritical region for $T_c^{(5)}$, $\T\simeq0.4$.

\begin{table}
	\caption{Critical point parameters for the first five critical points, $k=1,\ldots,5$: the critical temperatures $T_c^{(k)}$, reduced densities $\bar{n}_c^{(k)}$, pressures $P_c^{*(k)}$ and chemical potentials $\mu_{0c}^{(k)}$ at $f=1$ and $v^* = 1$.} \label{tab1}
	\begin{center}
		\begin{tabular}{|c|c|c|c|c|}
			\hline
			\multicolumn{5}{|c|}{Critical point} \\
			\hline
			$k$ &  $T_c^{(k)}$    &   $\bn_c^{(k)}$     &  $P_c^{*(k)}$  &  $\mu_{0c}^{(k)}$   \\
			\hline
			1 &		0.261408  &   0.535499   &    0.053526  &    0.517972    \\
			\hline
			2 &		0.284104  &   1.519144   &    0.268242  &    0.729561    \\
			\hline
			3 &		0.298771  &   2.512845   &    0.533432  &    0.867485   \\
			\hline
			4 &		0.309797  &   3.509555   &    0.827357  &    0.972564    \\
			\hline
			5 &		0.318686  &   4.507553   &    1.141058  &    1.058496    \\
			\hline
		\end{tabular}
		%\label{tab1}
	\end{center}
\end{table}

\section{Redefinition of the chemical potential}\label{REM}

Before proceeding to the extensive analytical calculations of the next section, we find it convenient to redefine the chemical potential by removing the annoying $\ln v$ term, additively shifting the value $p\mu$ in \eqref{Y}, \eqref{KK}, \eqref{Z} and other related places.

First of all, the appearance of the logarithm $\ln v$ with the \emph{dimensionful} argument $v$ is not mathematically clean: the arguments of logarithms have to be dimensionless $c$-numbers. This inconsistency is eliminated by taking into account the presence of de~Broglie thermal wavelength $\Lambda$ in the activity $\zeta$ (see \eqref{LA3}--\eqref{Lam}), as initiated in \cite{Retal25}. This results in the replacement of $v$ by the dimensionless ratio $v/\Lambda^3$ in GPF's integral representations as explicitly shown in \eqref{T}.

However, in theoretical calculations, it is crucial to work with most efficiently chosen variables and compact notations. Therefore, we define the new shifted dimensionless chemical potential $\mu$ via
\be\label{MEM}
\mu:=\mu_0+p^{-1}\ln\frac v{\Lambda^3}\,,
\ee
the combination that replaces the previously used dimensionless chemical potential $\mu_0 \equiv \mu_{\rm phys}/J_1$.

This redefinition does not influence any physical conclusions concerning the phase behavior of the CFM. For example, in Section~\ref{SWY} we shall see that the inverse critical temperature $p_c$ is determined from simultaneous vanishing of the second and third derivatives of the function $E(z)$ (cf. \cite[(28)]{KD22}), and is thus completely independent of the chemical potential. Moreover, with the definition \eqref{MEM}, our expressions will be identical to those used in previous work with the choice $v=1$, as in \cite{DSh24} and references quoted therein.

The set of dimensionless parameters $(p,\mu)$, with $\mu$ from \eqref{MEM}, is a convenient re-mapping of fundamental physical thermodynamic variables $(T,\mu_{\rm phys})$. Any particular pair $(p,\mu)$ uniquely defines the system's state just
as the pair $(T,\mu_{\rm phys})$ does. The mapping is invertible, which can be useful in calculations related to the entropy (see \cite[Sec. 7]{Retal25}, \cite{RDKPS26}).

\section{The strong-repulsion limit $J_2\gg J_1$}\label{J2B}

In this section, we consider the limiting case of strong repulsion interactions within the cells, $J_2\to\infty$, with finite attraction coupling $J_1>0$ (see Section~\ref{REX}).

Our starting point is the integral representation \eqref{Y}--\eqref{KK} with the replacement $v\mapsto v/\Lambda^3$ followed by the chemical potential shift \eqref{MEM}.
The large repulsion parameter $J_2$ enters through the quantity $r=\beta J_2$. Working with the temperature variable $p$, we write $r=p\,f$ and consider the large ratio of repulsion and attraction couplings $f\gg1$ (see \eqref{f}). Thus, we approximate the function $K(r;y+p\mu)$ from \eqref{KK} by retaining only the first non-trivial term in the sum:
\be\label{FUS}
K(r;y+p\mu)\simeq\sum_{n=0}^1\frac1{n!}\e^{(y+p\,\mu)n-\frac 12\,r n^2}=
1+\e^{y+p\,\mu-\frac12\,pf}+\mbox{exponentially small corrections}\,.
\ee
In the thermodynamic limit ${\mathcal N}\to\infty$, which allows us to
omit the subleading overall factor $\propto\sqrt {\mathcal N}$ in \eqref{Y},
the grand partition function takes the form
\be\label{BIF}
\Xi_{\mathcal N}(p,\mu;f\gg1)=\int_{-\infty}^\infty\!dy\e^{{\mathcal N}\,E(y)}\MM{with}
E(y)=\ln\left(1+\e^{y+p\,\mu-\frac12\,pf}\right)-\frac{y^2}{2p}\,.
\ee

This is the starting point for our analytical study of the CFM in the large-$f$ approximation. The analysis proceeds through the study of the extremal properties of the function $E(y)$ in \eqref{BIF}, beginning with the determination of the critical point.

\subsection{A simple way to the critical point}\label{SWY}

The aim of this section is to determine the critical point in the large-$f$ limit.
To simplify the following expressions we introduce short-hand notations
\be\label{SHN}
\e^{y-y_0}\equiv x>0\MM{and}y_0\equiv\frac p2\left(f-2\mu\right)\,,
\ee
and note that $\displaystyle{\frac{dx}{dy}=\frac{d\e^{y-y_0}}{dy}=\e^{y-y_0}=x}$.
To locate the critical point, we analyze the standard Landau-theory conditions
$E'(y)=E''(y)=E'''(y)=0$ and $E^{\rm iv}(y)<0$.

The needed derivatives of $E(y)$ from \eqref{BIF} are:
\be\label{POE}
E'(y)=\frac{\e^{y-y_0}}{1+\e^{y-y_0}}-\frac yp=\frac{p\,x-y-yx}{p(1+x)}\,,
\ee
\be\label{POD}
E''(y)=\frac{\e^{y-y_0}}{(1+\e^{y-y_0})^2}-\frac1p=
-\,\frac{x^2+(2-p)\,x+1}{p(1+x)^2}\,,
\ee
\be\label{POT}%=\e^{y-y_0}\,\frac{1-\e^{y-y_0}}{(1+\e^{y-y_0})^3}
E'''(y)=\frac{x(1-x)}{(1+x)^3}\,, \MM{and}
E^{\rm iv}(y)=x\,\frac{(1-x)^2-2\,x}{(1+x)^4}\,.
\ee
Apart from ubiquitous dependence on $x$, the function $E'(y)$ contains explicit dependencies on $y$ and $p$, while $E''(y)$ depends explicitly only on $p$.
No such explicit dependencies on $y$ and $p$ appear in $E'''(y)$.
The condition $E'''(y)=0$ immediately yields the unique positive solution $x=1$,
which implies $y=y_0$.

At this point ($x=1$, $y=y_0$), the second derivative becomes
\be\label{POF}
E''(y_0)=\frac{p-4}{4p}\,.
\ee
This means that $E''(y_0)<0$ for any $y_0$ if $p<4$, and any extremum of $E(y)$ appears to be a simple maximum in this region of $p$.

At the critical value of $p$, $p_c=4$, the function $E''(y_0)$ vanishes signalling a more complicated behavior of $E(y)$. Its maximum becomes degenerate with
$E''(y_0)=E'''(y_0)=0$ and $E^{\rm iv}(y_0)=-\frac18<0$. The negative sign of the fourth derivative ensures the stability of the critical point.

With $x=1$, the extremum condition $E'(y_0)=0$ yields
\be\label{POC}
p-2y_0=0\,.
\ee
At $p=p_c=4$, we hence obtain $y_c\equiv y_0|_{p=p_c}=p_c/2=2$ for the maximum position of $E(y)$ at the critical point. This fixes, through \eqref{SHN}, the critical value of the chemical potential:
\be\label{MUC}
\mu=\mu_c=\frac12(f-1)\,.
\ee

Thus, in the strong-repulsion limit, we find a critical point at an inverse critical temperature $p_c=4$ and a critical chemical potential $\mu_c = (f-1)/2$.
With these parameters, the function $E(y)$ from \eqref{BIF} becomes
%(this is a special case of \eqref{MIP} at $p=4$)
\be\label{ECY}
E_c(y)=\ln\left(1+\e^{y-2}\right)-\frac{y^2}8\,,
\ee
which has a flat degenerate maximum at $y=y_c=2$.
The maximum value of this function is
\be\label{MVV}
E_c(y_c)=E_c(2)=\ln2-\frac12\simeq0.193147\,,
\ee
which agrees with previous numerical work.
In fact, the maximum value $E_c(y_c)$ from \eqref{MVV} represents the \emph{exact analytic expression} for numerical values $0.1931...$ given in \cite[Table 2]{DKPP26} for the dimensionless pressure $\tilde P_c^{(1)}$ at $f=5$ and $f=10$, as well as for the critical pressure in \cite[Fig.~1]{PS15}.

%According to \cite[(17)]{DKPP26}, this quantity actually coincides with our dimensionless combination $\beta Pv=pP^*$ appearing in \eqref{eod}, \eqref{eons}, and \eqref{ERR}.

Another important critical parameter is the mean cell occupancy (reduced particle density) $\bar n_c=y_c/p_c=1/2$.
This value, along with $p_c=4$, appears to be quite universal and agrees with numerical findings in \cite[Table 1]{KKD20}, \cite[p. 249]{KKD18}, and
\cite[Table 2]{Petal25} over a wide range of the parameter $f\gtrsim3$.
Moreover, these values do not deviate far away even in the opposite marginal case $f=1$, for which we obtain $p_c(f{=}1)\simeq3.825$ and $\bar n_c(f{=}1)\simeq0.535$ (see Section~\ref{PDf1}, Table \ref{tab1}).

\subsection{Temperature dependence of $E(y)$ at $\mu=\mu_c$}\label{SPE}

\begin{figure}[htbp]
	\centering
	\includegraphics[width=0.6\textwidth]{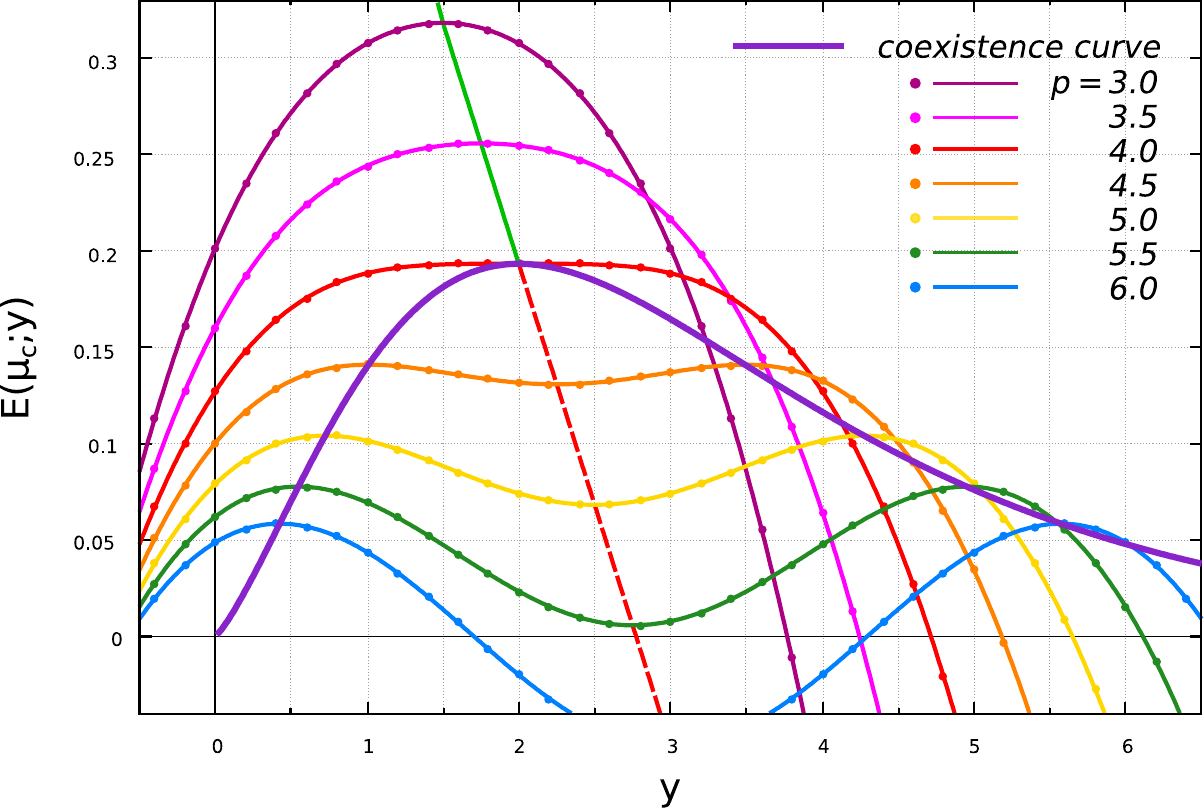}
	\caption{Plots of the function $E(\mu_c;y)$ from \eqref{BIP} (thin solid curves) and $E(y)$ at $\mu_0 = \mu_c$ from \eqref{EE} (dots) at seven values of the parameter $p$ listed in the legend.
    For $p<p_c=4$, there is a simple single maximum. At $p=p_c$, the maximum becomes
    flat, corresponding to the critical point, with its height $E_c(y_c)=\ln2-1/2$
    from \eqref{MVV}.
    For $p>p_c$, the central extremum smoothly becomes a minimum and two new
    symmetric maxima appear, signaling a two-phase coexistence.
    The plots are symmetric with respect to their central extremum
    positions $\bar{y}^{(1)}=p/2$. The straight inclined line given by \eqref{clin} connects these extrema at $\bar{y}^{(1)}=p/2$ for all curves.
    The thick purple plot running through the side maxima shows the coexistence (binodal) curve $E_{\rm max}(\mu_c;y)$ from \eqref{eMAX}.
   }
	\label{f3}
\end{figure}

At the critical value of the chemical potential, $\mu=\mu_c=(f-1)/2$ (see \eqref{MUC}), the function $E(y)$ from \eqref{BIF} simplifies to
\be\label{BIP}
E(\mu_c;y)=\ln\left(1+\e^{y-p/2}\right)-\frac{y^2}{2p}\,.
\ee
The dependence on the large parameter $f$ has disappeared,
and the sole remaining physical parameter in \eqref{BIP} is the inverse temperature $p$. The shape of the function $E(\mu_c;y)$ changes as $p$ crosses the critical value $p_c=4$, as shown in Fig.~\ref{f3}.

The plot of $E(\mu_c;y)$ at $p=6$ is analogous to that given in \cite[Fig.~4]{KKD20}. Both are very similar even in numerical values, though the latter was obtained using the complete infinite sum \eqref{R} rather than the truncated approximation \eqref{FUS} and a different value of critical chemical potential $\mu_c$.

Fig.~\ref{f3} also demonstrates perfect agreement between the plots of the function $E(\mu_c;y)$  from Eq.~\eqref{BIP}, shown by thin solid curves and obtained within the truncated approximation~\eqref{FUS}, and the plots of the function $E(y)$ at $\mu_0 = \mu_c$ from Eq.~\eqref{EE}, shown by solid dots and calculated numerically using the complete infinite sum~\eqref{R}. While the truncated approximation results in a single phase transition, the complete infinite sum treatment at $f=10$ allows for the emergence of multiple first-order phase transitions and, consequently, multiple critical points, as demonstrated in~\cite[Table~2]{DKPP26} and~\cite[Table~2]{Petal25}. Fig.~\ref{f3} span values of $y$ and $E(y)$ corresponding to the first phase transition in the sequence. The second and third transitions in the sequence occur at pressures that are approximately 190 and 570 times higher than that of the first phase transition, respectively. By contrast, the phase diagrams for the marginal case $f=1$, shown in Fig.~\ref{f1} (Sec.~\ref{PDf1}), indicate that the corresponding pressure differences is at most 5 and 10 times, respectively.

Vanishing of $E'(\mu_c;y)$ yields the extremum condition
\be\label{EXC}
\e^{y-p/2}=\frac y{p-y}\,,
\ee
identifying the extremum positions $\bar y$ as functions of temperature $p$.
An evident solution to this equation is $\bar y^{(1)}=p/2$, the same as that
following from \eqref{POC}. As long as $p<p_c$, the coordinate $\bar y^{(1)}=p/2$ corresponds to stable simple maxima of $E(\mu_c;y)$.
At $p=p_c=4$, the maximum becomes degenerate and acquires the flat shape of the red curve.
In the region $p>p_c$, the extremum at $y=\bar y^{(1)}$ becomes a minimum, and
two side maxima of equal height appear, whose positions are symmetric with respect to $\bar y^{(1)}=p/2$ and given by two further solutions to the extremum condition \eqref{EXC}, $\bar y^{(2)}<\bar y^{(1)}$ and $\bar y^{(3)}>\bar y^{(1)}$.
This bifurcation corresponds to the emergence of a two-phase coexistence, with the two maxima representing the coexisting low-density and high-density phases, and thus being the binodal points. The minima at $\bar y^{(1)}=p/2$ with $p>4$ do not correspond to stable physical states. These are the points belonging to the van der Waals loops, as we shall see in Section~\ref{VWL}, Fig.~\ref{f5}.

The heights of extrema corresponding to the solution $\bar y^{(1)}$ are given by
\be\label{MIP}
E(\mu_c;p/2)=\ln2-\frac p8\,,
\ee
which reduces to the previous result \eqref{MVV} at $p=p_c=4$. The straight line joining these extrema at different values of $p$, is given by
\be\label{clin}
E^{(1)}(\mu_c;y)=\ln2-y/4\,,
\ee
which is obtained from \eqref{BIP} at $p=2y$.

Our next task is to identify the two remaining solutions to \eqref{EXC}.
A simple rearrangement of this equation yields
\be\label{EXR}
\e^{y/2+(y-p)/2}=-\frac2{y-p}\,\cdot\,\frac y2\,,\MM{whence}
\frac{y-p}2\,\e^{(y-p)/2}=-\frac y2\,\e^{-y/2}\,.
\ee
The last equation has the form $W{\rm e}^W=z$, which allows us to obtain its solution in terms of
the Lambert $W$-function $W(z)$ \cite{Lambert}. Hence we infer
\be\label{GES}
p=y-2W\Big(-\frac{y}2\,\e^{-y/2}\Big),
\ee
relating the parameter $p$ to corresponding maximum positions $\bar y_{\rm max}$ in the region $p\ge p_c$.

At this point, it is useful to introduce two short-hand notations, which will allow us to express the desired result in a compact and elegant form. These are:
\be\label{XZ}
X=X(y)\equiv-\frac{y}2\,\e^{-y/2}\MM{and}Z_\nu=Z_\nu(y)\equiv\frac y{2W_\nu(X)}\,.
\ee
The index $\nu=-1,0$ indicates the relevant branches of the Lambert function $W_\nu$.
Our special value $y=2$ is precisely the branching point of the Lambert function $W(X)$ with negative argument $X$: in the interval between $X(2)=-\e^{-1}$ and zero, this function is double-valued with two real branches $W_{-1}(X)$ and  $W_0(X)$. In terms of these, the curves running through the maxima of $E(\mu_c;y)$ are given by the functions
\be
E_\nu(y)=\ln(1-Z_\nu)-\frac y{2(1-Z_\nu^{-1})}\,,\qquad\qquad\nu=-1,0,
\ee
with $Z_\nu$ defined in \eqref{XZ}. The desired result for the coexistence curve is thus
\be\label{eMAX}
E_{\rm max}(\mu_c;y)=\begin{cases}
E_{-1}(y),\;\;0\le y\le2\\
E_0(y),\qquad y\ge2\,.
\end{cases}
\ee
The function $E_{\rm max}(\mu_c;y)$ is given in Fig.~\ref{f3} by the thick purple curve.

The straight line $E^{(1)}(\mu_c;y)$ from \eqref{clin} passing through central extrema
\be\label{cex}
\bar y^{(1)}=\begin{cases}
\bar y_{\rm max},\;\;p\le p_c\\
\bar y_{\rm min},\,\;\;p>p_c
\end{cases}
\ee
of $E(\mu_c;y)$, is also related to above branches of the Lambert function $W(X)$ via
\be
E^{(1)}(\mu_c;y)=\begin{cases}
E_0(y),\;\;0\le y\le2\\
E_{-1}(y),\;\;y\ge2.
\end{cases}
\ee

Due to the definition of the equation of state $pP^*=E(\bar y_{\rm max})$ in \eqref{ERR} and the identification of the cell occupancy $\bn$ as $\bn=\bar y_{\rm max}/p$ in \eqref{ncy}, we see that the function $E_{\rm max}(\mu_c;y)$ from \eqref{eMAX} actually represents the equation of state in the form
\be\label{BNM}
pP^*=E_{\rm max}(\mu_c;p\,\bn)\,.
\ee
Up to the axes' scales through the factor $p$, this is the pressure-density equation of state of CFM at the critical chemical potential $\mu_c=(f-1)/2$ in the large-$f$ approximation.

\subsubsection{A note on symmetry}\label{ANOT}

To confirm the apparent symmetry of the curves in Fig.~\ref{f3} with respect to $y=p/2$, we shift in $E(\mu_c;y)$ the variable $y$
via $y-p/2=s$. Thus we obtain the function
\be\label{BIS}
\hat E(\mu_c;s)=\ln\left(1+\e^s\right)-\frac1{2p}\,\Big(s+\frac p2\Big)^2,
\ee
symmetric with respect to the ordinate axis.
To see that $\hat E(\mu_c;s)$ is an even function, we change $s\mapsto-s$
in \eqref{BIS} and obtain
\begin{align}\nn%\label{BIW}
&
\hat E(\mu_c;-s)=\ln\left(1+\e^{-s}\right)-\frac1{2p}\,\Big(-s+\frac p2\Big)^2=
\ln\left[\e^{-s}(1+\e^s)\right]-\frac1{2p}\,\Big(s-\frac p2\Big)^2=
\\\nn&
=\ln\left(1+\e^s\right)-\frac1{2p}\Big[\Big(s-\frac p2\Big)^2+2ps\Big]
=\ln\left(1+\e^s\right)-\frac1{2p}\,\Big(s+\frac p2\Big)^2=\hat E(\mu_c;s)\,.
\end{align}

For the function $\hat E(\mu_c;s)$, all curves corresponding to that in Fig.~\ref{f3} become symmetric with respect to the axis $s=0$.
The central extremum positions of $E(\mu_c;y)$ at $\bar y^{(1)}=p/2$ are mapped to a single point $s=0$. The heights of these extrema are given by \eqref{MIP}.

\begin{figure}[htbp]
	\centering
	\includegraphics[width=0.6\textwidth]{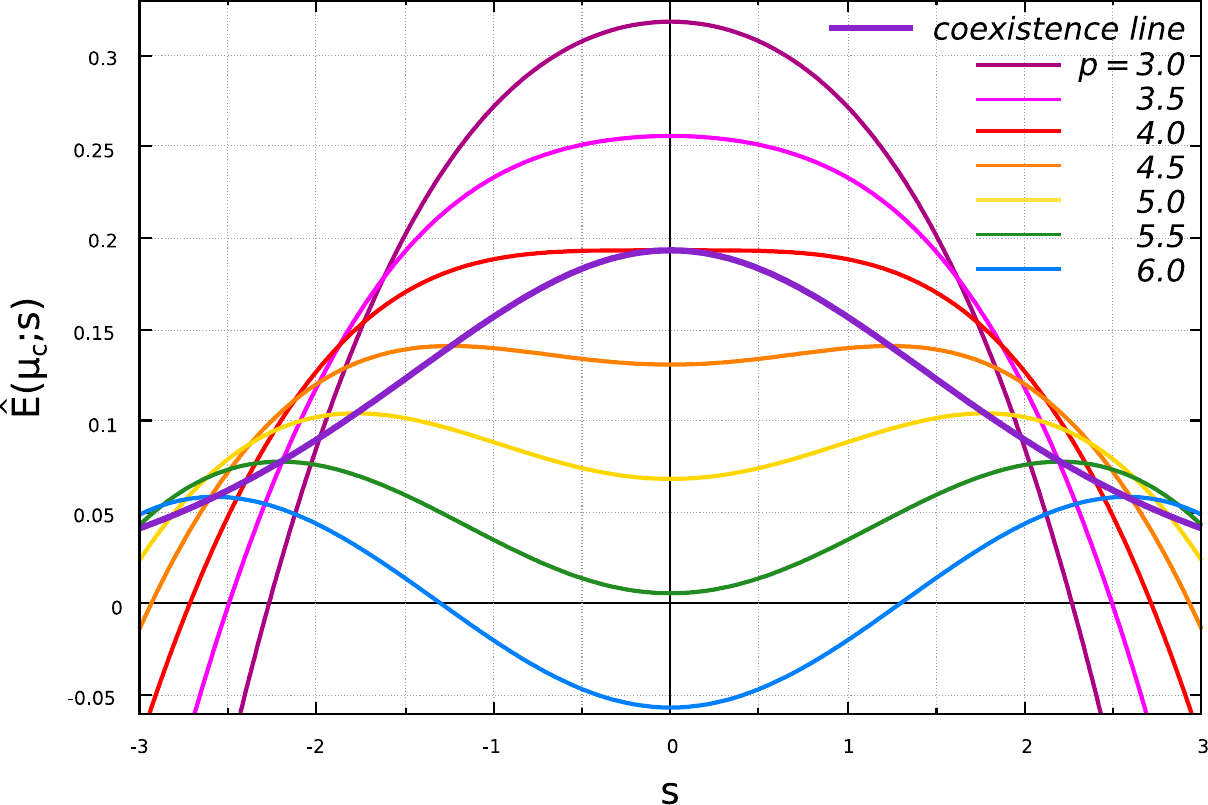}
	\caption{Plots of the function $\hat E(\mu_c;s)$ from \eqref{BIS} at the
    same values of $p$ as in Fig.~\ref{f3}. All curves are symmetric with respect to the $s=0$ axis as expected for the even function $\hat E(\mu_c;s)$.
	}
	\label{f4}
\end{figure}

For $\hat E(\mu_c;s)$, the extremum condition analogous to \eqref{EXC} is given by
\be\label{EXS}
\e^s=\frac{p+2s}{p-2s}\,.
\ee
It is easy to see that it is trivially fulfilled at $s=0$. For the symmetric side maxima, this condition yields
\be\label{DDN}
p=2s\,\frac{\e^s+1}{\e^s-1}=2s\coth\frac s2\,,
\ee
which is the counterpart of the relation \eqref{GES}. Notably, \eqref{DDN} involves only  elementary functions, in contrast to \eqref{GES}, expressed in terms of the Lambert $W$-
function. Substituting \eqref{DDN} into \eqref{BIS}, we obtain the following simple result for the function tracing the maxima of $\hat E(\mu_c;s)$ in the present symmetric setting:
\be\label{ICS}
\hat E_{\rm max}(\mu_c;s)=\ln\left(1+\e^s\right)+\frac s{\e^{-2s}-1}\,.
\ee
Due to an appropriate symmetrization of the problem, this result is much simpler than its counterpart obtained in \eqref{eMAX}.
\bigskip

Finally, expanding the function $\hat E(\mu_c;s)$ in powers of $s$ around the origin we obtain
\be\label{LAE}
\hat E(\mu_c;s)\simeq\hat E(\mu_c;0)+\frac{p-4}{8p}\,s^2-\frac{s^4}{192}+O(s^6)
\ee
with the constant term $E(\mu_c;0)$ given in \eqref{MIP}.

The functional form appearing in \eqref{LAE} is that of the usual Landau expansion for a second-order phase transition.
The coefficient of the $s^2$ term changes its sign at the critical value $p_c=4$.
The thermodynamic stability is ensured by the negative sign of the $s^4$ term, and only even powers of $s$ appear in \eqref{LAE}.
Thus, in the strong-repulsion limit and at the critical value of the chemical potential, the CFM has a Landau expansion, which is equivalent in form and symmetry to that of the classical lattice gas or other Ising-like systems in zero external field.
However, since the CFM is mean-field by construction, it lacks the spatial fluctuations necessary to produce the non-trivial exponents characteristic of these systems.
Due to this inherent restriction, the CFM can be said to belong to the mean-field Ising universality class.
\bigskip

In the following section, we shall consider the behavior of the function $E(y)$ from \eqref{BIF} for chemical potentials $\mu$ differing from its critical value $\mu_c$.

\subsection{Temperature dependence of $E(y)$ with $\mu\ne\mu_c$}\label{TDK}

In this section, we relax the constraint $\mu=\mu_c$ and study the special points
of the function $E(y)$ from \eqref{BIF} for independently varying temperature
$p$ and chemical potential $\mu$.

\subsubsection{Horizontal inflection points}\label{HIP}

We start by studying the special points where the first and the second derivatives of the function $E(y)$ from \eqref{BIF} vanish simultaneously. As we shall see, the condition $E'(y)=E''(y)=0$ has a special significance, since it defines the spinodal points lying on the boundaries of thermodynamically metastable and unstable regions for $p>p_c$.
%\bigskip

The first derivative of the function $E(y)$ from \eqref{BIF} is given in \eqref{POE}. It becomes zero when $p\,x-y-yx=0$, and hence the extremum condition for this function is (cf. \eqref{EXC})
\be\label{C1}
y=p\,\frac x{1+x}\,.
\ee
The second derivative $E''(y)$ from \eqref{POD} vanishes when $x^2+(2-p)\,x+1=0$,
and this quadratic equation has the solutions
\be\label{EEE}
x_{1,2}(p)=\frac p2-1\pm\sqrt{p\,\Big(\frac p4-1\Big)}\,.
\ee
We shall stick to the convention that the index "$1$" refers to the plus sign before the radical.

When $p<p_c=4$, there are no real solutions $x_{1,2}$, and $E''(y)$ is always negative as it should be for thermodynamically stable homogeneous phases corresponding to single maxima of $E(y)$.
At $p=p_c=4$, the first real solution appears, $x_0=1$. This specific case has been studied in Section \ref{SWY}.

In the region $p>4$ we obtain two different solutions from \eqref{EEE}. The derivatives $E'(y)$ and $E''(y)$ vanish simultaneously when we combine the conditions \eqref{C1}
and \eqref{EEE} into
\be\label{Y12}
\tilde y_{1,2}(p)=p\,\frac{x_{1,2}(p)}{1+x_{1,2}(p)}\,,
\ee
which defines the coordinates $\tilde y_1(p)$ and $\tilde y_2(p)$ of new special points.

By definitions in \eqref{SHN}, vanishing of $E''(y)$ at $x=x_{1,2}(p)$ implies that
\be\label{YRT}
\e^{y-y_0}=x_{1,2}(p),\mm{that is,}y-y_0=\ln x_{1,2}(p),\mm{and thus,}
y+p\,\mu-\frac12\,pf=\ln x_{1,2}(p)\,.
\ee

A substitution of \eqref{Y12} into the last equation in \eqref{YRT} yields the values of $\mu$, at which the present setting is possible, namely
\be\label{M12}
\mu_{1,2}(p)=\frac12\,f-\frac{x_{1,2}(p)}{1+x_{1,2}(p)}+
\frac1p\,\ln x_{1,2}(p)\,.
\ee
For such $\mu_1$ and $\mu_2$ with any $p>4$, we have
$E'(\tilde y_{1,2})=E''(\tilde y_{1,2})=0$, and the function $E(y)$ has horizontal
inflection points with $E'''(\tilde y_1)<0$, $E'''(\tilde y_2)>0$, and
$|E'''(\tilde y_1)|=E'''(\tilde y_2)$ (see \eqref{POT} and Fig.~\ref{f5}).
Explicitly, we have the symmetric expressions
\be\label{oas}
E'''(\tilde y_{1,2})=\mp p^{-2}\sqrt{p(p-4)}\,,\qquad p\ge4\,.
\ee

The chemical potentials $\mu_1(p)$ and $\mu_2(p)$ define the spinodal points for each given temperature $p$, separating thermodynamically unstable regions from metastable ones --- see Fig.~\ref{f5}.

\subsubsection{Spinodal symmetry}\label{EAA}

A numerical check performed at $p=6$ and $f=10$ showed that chemical potentials $\mu_1$ and $\mu_2$ from \eqref{M12}, related to horizontal inflection points of $E(y)$,
obey the inequality
\be
\mu_1<\mu_c<\mu_2\MM{with}\mu_c=\frac12(f-1)\,.
\ee
Therefore we write these values as
\be
\mu_1(p)=\mu_c-\Delta\mu_1(p)\MM{and}\mu_2(p)=\mu_c+\Delta\mu_2(p)
\ee
with \emph{positive} deviations $\Delta\mu_1$ and $\Delta\mu_2$
of chemical potentials $\mu_1$ and $\mu_2$ from the central value $\mu_c$.
The difference of these deviations is given by
\be
\Delta\mu_1(p)-\Delta\mu_2(p)=-1+\frac{x_1(p)}{1+x_1(p)}+\frac{x_2(p)}{1+x_2(p)}-
\frac1p\,\ln\Big(x_1(p)\,x_2(p)\Big)=0\,.
\ee
This combination vanishes because $x_1(p)\,x_2(p)=1$, which follows directly from the definition of $x_{1,2}(p)$ in \eqref{EEE}. Thus,
\emph{the values $\mu_1(p)$ and $\mu_2(p)$ from \eqref{M12} are symmetric with respect to the critical value of chemical potential $\mu_c$}. That is,
\be\label{SPS}
\mu_1(p)+\mu_2(p)=2\mu_c\,,\qquad\text{or equivalently,}\qquad
\Delta\mu_1(p)=\Delta\mu_2(p)\,,
\ee
for any $p$ in the range $p>4$.

This symmetry means that in the present approximation of large $f$, the spinodal stability limits of the high- and low-density phases are perfectly symmetric with respect to the critical chemical potential $\mu_c$.

Moreover, a similar symmetry holds for inflection-point coordinates $\tilde y_1(p)$
and $\tilde y_2(p)$ from \eqref{Y12}. This equation, along with \eqref{EEE} yields
\be
\tilde y_1(p)+\tilde y_2(p)=p\,.
\ee
This means that the spinodal points $\tilde y_1(p)$ and $\tilde y_2(p)$ are positioned symmetrically with respect to $\bar y^{(1)} = p/2$, just as the coexistence maxima in Section \ref{SPE}. This common center of symmetry is fundamental for the large-$f$ limit of the CFM, but it is expected that the same symmetry properties do not hold for the full model from Section \ref{RER} with arbitrary $f$.

\subsubsection{An example}

Let us give an explicit numerical example by taking $p=6$, which is the same choice as in figures 3~a) and 3~b) in \cite[p.~13]{KKD20}.

At $p=6$, we have $x_{1,2}=2\pm\sqrt3$\,, the coordinates of inflection points are $\tilde y_{1,2}=3\pm\sqrt3$\,, and the associated values of the chemical potential are
\be\label{MMM}
\mu_{1,2}=\frac12\,f-\frac16\,(3\pm\sqrt3)+\frac16\,\ln\!\big(2\pm\sqrt3\,\big)\,.
\ee
With $\mu=\mu_c=\frac12(f-1)$, we have
\begin{align}%\label{BIV}
&
\mu_1=\mu_c-\frac16\,\Big[\sqrt3-\,\ln\!\big(2+\sqrt3\,\big)\Big],
\\&%\label{BIW}
\mu_2=\mu_c+\frac16\,\Big[\sqrt3+\,\ln\!\big(2-\sqrt3\,\big)\Big]\,.
\end{align}
It follows immediately that $\mu_1+\mu_2=2\mu_c$ and $\Delta\mu_1=\Delta\mu_2$ because $\ln\!\big[\big(2+\sqrt3\big)\big(2-\sqrt3\big)\big]=\ln(4-3)=0$.

We also observe the symmetry of the inflection points $\tilde y_{1,2}=3\pm\sqrt3$\,, with respect to the central value $p/2=3$ at $p=6$, as discussed at the end of the preceding section.

\subsubsection{The spinodal curve}\label{ESS}

The spinodal curve joins all inflection points of the function $E(y)$ from \eqref{BIF} in the region $p>4$.

\begin{figure}[htbp]
	\centering
	\includegraphics[width=0.8\textwidth]{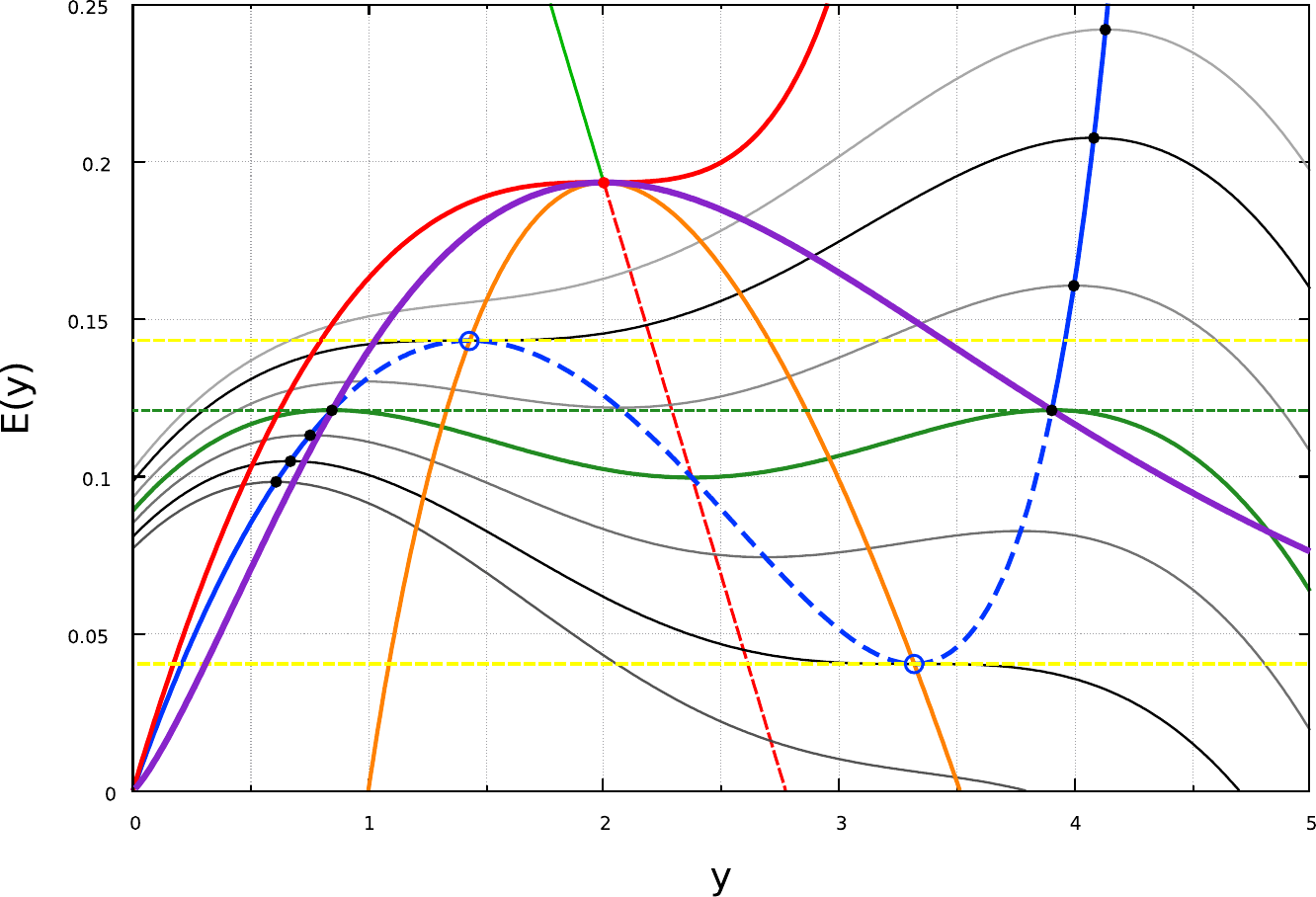}
	\caption{The function $E(y)$ from \eqref{BIF} at $p=4.75$ and $f=10$, for seven values of $\mu$ growing from $4.468$ to $4.53$. Gray curves show the full family, while selected ones are highlighted for $\mu=\mu_1\simeq4.478$ and $\mu=\mu_2\simeq4.522$ (black curves with inflection points marked by blue circles), and $\mu=\mu_c=4.5$ (dark-green curve, with its two equal-height maxima connected by the green dashed horizontal line). The thick blue curve is the isotherm $E_{\rm ext}(4.75;y)$ from
\eqref{SDS}, with its stable branches (marked by black dots) above the binodal. The thick red curve is the critical isotherm $E_{\rm ext}(4;y)$, passing through the critical point (red dot) at $y=2$. The thick purple curve is the binodal $E_{\rm max}(\mu_c;y)$ from \eqref{eMAX}, and the thick orange curve is the spinodal $E_{\rm inf}(y)$ from \eqref{SPI}. The straight inclined light-green line given by $E^{(1)}(\mu_c;y)$ from \eqref{clin} shows the isotherm values at $\mu=\mu_c$ for stable states at $p<p_c$ (cf.\ Fig.~\ref{f3}). Its continuation to the forbidden unstable region with $y>2$ shown as a red dashed line passes through the minima of $E(\mu_c,p;y)$ at $y=\bar y^{(1)}=p/2$, as in Fig.~\ref{f3}. Yellow dashed lines indicate the horizontal tangents of inflection points corresponding to $\mu=\mu_1\simeq4.478$ and $\mu=\mu_2\simeq4.522$. These inflection points coincide with the extrema of the van der Waals loop at $\tilde y_1\simeq3.319$ and $\tilde y_2\simeq1.431$ (blue circles) crossed by the spinodal curve.}
	\label{f5}
\end{figure}

To identify this curve, we first note that with notations \eqref{SHN}, the function  $E(y)$ is given by
\be\label{BIX}
E(y)=\ln(1+x)-\frac{y^2}{2p}\,.
\ee
As discussed in Section \ref{HIP}, the spinodal inflection points are identified by the double condition $E'(y)=E''(y)=0$, where the derivatives of the function $E$ are given in \eqref{POE} and \eqref{POD}. Starting from these equations, we write the relevant condition as
\be\label{SIX}
\begin{cases}
px-y(1+x)=0\\
(1+x)^2-px=0\,.
\end{cases}
\ee
From the second equation in \eqref{SIX}, we have $px=(1+x)^2$. Substituting this into the first equation we obtain $1+x=y$, and this is what we need for the argument of the logarithm in \eqref{BIX}. With this identification, we derive $p$ as the function of $y$ via $p=(1+x)^2/x=y^2/(y-1)$. This allows us to express the second term in \eqref{BIX} in terms of the variable $y$ satisfying the system of equations \eqref{SIX}. Thus we obtain for the spinodal curve
\be\label{SPI}
E_{\rm inf}(y)=\ln y+\frac{1-y}2\,.
\ee
This curve runs through the inflection points of $E(y)$, as shown in Fig.~\ref{f5}, for all temperatures $p$ in the region $p>4$.
At $y=2$, the curve $E_{\rm inf}(y)$ attains its maximum coinciding with the degenerate maximum of the function $E_c(y)$ at the critical point, with the value $E_{\rm inf}(2)=E_c(2)=\ln2-1/2$ in agreement with \eqref{MVV}.

\subsubsection{Isotherms and van der Waals loops}\label{VWL}

In Section \ref{SPE}, where we fixed $\mu$ at its critical value $\mu_c$, for each chosen $p>p_c$ we have encountered only two points of corresponding isotherms, situated at equal maxima of the function $E(\mu_c,p;y)$ and physically related to points of phase coexistence.
In order to construct the full isotherms for the function $E(\mu,p;y)$ and obtain the related van der Waals loops, we fix the temperature $p$ and allow the chemical potential $\mu$ to vary.

The process is illustrated in Fig.~\ref{f5}. For a chosen temperature $p>p_c=4$, we look at the family of dependencies of $E(\mu,p;y)$ on $y$ for different values of the chemical potential $\mu$. With lowest values of $\mu$, the function $E(y)$ has its maximum rather close to zero, and monotonically decreases as $y$ grows further. At $\mu=\mu_1(p)$, as discussed in Section~\ref{HIP}, the curve $E(y)$ has an inflection point with horizontal tangent. As soon as $\mu$ exceeds $\mu_1$, a lower-height side maximum appears on the right-hand side, which can be attributed to possible appearance of metastable states. When the value of $\mu$ eventually reaches its critical value $\mu_c$, the two maxima occurring at smaller and larger $y$ attain the same height as shown by the dark-green curve in Fig.~\ref{f5}. This signals a first-order phase transition in the system, and the maximum points belong to the phase-coexistence curve for the initially chosen temperature $p$. When $\mu>\mu_c$, the side maxima on the right "win" and represent physically stable states in this region of the chemical potential.

A similar family of curves $E(\mu,p;y)$ can be drawn for each other temperature $p>p_c$.
As $p$ approaches $p_c$, the coexistence points become closer and closer to each other, and finally they merge into a single point shown by the red dot in Fig.~\ref{f5} at $p=p_c$.

To identify the isotherms of $E(\mu,p;y)$ as functions of $y$, we recall that
at any $p$ and $\mu$, the extremum condition $E'(y)=0$ is given by $px-y-yx=0$ (see \eqref{POE}), which yields
\be\label{XEX}
x=\frac y{p-y}\,.
\ee
Substituting this solution into $E(y)$ from \eqref{BIF} we obtain the function
\be\label{SDS}
E_{\rm ext}(p;y)=\ln\,\frac{p}{p-y}-\frac{y^2}{2p}\,.
\ee
For any given temperature $p$, this function passes through all extrema of $E(\mu,p;y)$, the most relevant of which correspond to stable thermodynamic states when these extrema are the global maxima.

In the special case of $p=p_c$ and varying $\mu$, the \emph{critical isotherm} passes through the critical point at $y=2$ and has a horizontal inflection point there. The critical isotherm $E_{\rm ext}(4;y)$ is shown by the red curve in Fig.~\ref{f5}.

All isotherms with $p<p_c$ (not shown in the figure) are monotonically increasing functions running above this special curve. Their values at $\mu=\mu_c$ are given by the light-green line on the top of Fig.~\ref{f5} (cf. Fig.~\ref{f3}.).

When $p>p_c$, the isotherms $E_{\rm ext}(p;y)$ have the form of van der Waals loops with maxima and minima located at $\tilde y_1(p)$ and $\tilde y_2(p)$ defined in \eqref{Y12}. In the special case of $p=4.75$, the isotherm $E_{\rm ext}(4.75;y)$ is given by a thick blue curve in Fig.~\ref{f5}.

The intersections of functions $E_{\rm ext}(p;y)$ with coexistence maxima of $E(\mu,p;y)$ at $\mu=\mu_c$ form the binodal curve given by the thick purple plot in Fig.~\ref{f5}.
Only the branches of isotherms lying above the binodal correspond to physically stable states. These branches of the isotherm $E_{\rm ext}(4.75;y)$ in Fig.~\ref{f5} are marked by black points.

The spinodal curve \eqref{SPI} defined in the preceding section, goes through the inflection points of $E(y)$ for $p>4$, and thus crosses all maxima and minima of corresponding van der Waals loops.

The red dotted inclined line shows the minima of the function $E(\mu_c,p;y)$ for any $p>4$, as in Fig.~\ref{f3}. At the same time, this line indicates the inflection points of the van der Waals loops in the middle of the instability region.

\subsubsection{A Landau expansion for $\mu\ne\mu_c$}

Finally, we extend the Landau expansion mentioned at the end of Section~\ref{ANOT} to the general case $\mu\ne\mu_c$.

By analogy with \eqref{BIS}, we shift in \eqref{BIF} the variable $y$ via $y=s+pf/2-p\mu$ to obtain
\be\label{MIS}
\hat E(\mu;s)=\ln\left(1+\e^s\right)
-\frac1{2p}\,\Big(s+p\Big(\frac12+\mu_c-\mu\Big)\Big)^2\,,
\ee
where we took into account the definition of $\mu_c$ in \eqref{MUC}.
Expanding the function $\hat E(\mu;s)$ in powers of $s$ in analogy with \eqref{LAE}, we obtain the Landau expansion
\be\label{MIC}
\hat E(\mu;s)=\ln2-\frac p2\Big(\mu-\mu_c-\frac12\Big)^2
+(\mu-\mu_c)\,s+\,\frac{p-4}{8p}\,s^2-\frac{s^4}{192}+O(s^6).
\ee
At $\mu=\mu_c$, the linear term in $s$ vanishes, and the expansion \eqref{MIC} reduces to the even function obtained in \eqref{LAE}.

As at the end of Section~\ref{SPE}, we observe an equivalence between the CFM in the strong-repulsion limit and classical lattice gases with non-zero chemical potential or the Ising model in the presence of an external ordering field, considered within the framework of the mean-field (or Landau) approximation.

\subsection{Results in terms of the reduced density $\bar n$}\label{ph}

Until now, we have analyzed the values of the function $E$ at its global maximum positions $\bym$. In view of the relations \eqref{ERR} and \eqref{ncy}, this corresponds to expressing the equation of state as $pP^*=E_{\rm max}(p\,\bn)$, which differs from the conventional thermodynamic form $P^*=P^*(\T,\bar n)$.
Although these two representations differ only by scaling of the basic thermodynamic variables $P^*$ and $\bn$ by the inverse temperature $p$, it is not evident how the final analytical results of preceding sections should be modified to provide a transition to the desired standard form $P^*=P^*(\T,\bar n)$.
This situation is reminiscent of the existence of two distinct analytical expressions \eqref{eMAX} and \eqref{ICS} representing essentially the same physical binodal curve. These are expressed in terms of different variables $y$ and $s$; one involves two non-trivial branches of the transcendental Lambert function, while the other is expressed via elementary exponential and logarithmic functions. The key point was that different equations --- \eqref{EXC} and \eqref{EXS} had to be solved to produce these two versions of the binodal curve.

With the relevant information from preceding sections at hand, we can efficiently achieve the goal of producing the standard "physical" equation of state $P^*=P^*(\T,\bar n)$ by changing the integration variable in \eqref{BIF} via
\be\label{YTU}
u:=\frac y p
\ee
and apply the Laplace method once again to the transformed integral.
It is essential that the variable change in \eqref{YTU} gives us direct access to the reduced density $\bn$ via
\be\label{TT}
\bum=\bn\,.
\ee
This can be explicitly verified by reproducing the argument of Section \ref{RER} that led to the relation $\bym=p\,\bn$ in \eqref{ncy}.

Introducing a counterpart of the function $E(y)$ given in \eqref{BIF} and using the notation $x$ by analogy to \eqref{SHN}, we start with
\be\label{SOS}
E(p,\mu;u)=\ln(1+x)-\frac12\,p\,u^2,\MM{where}x:=\e^{p(u+\mu-\frac12\,f)}\,.
\ee
The derivatives of $E(u)$ with respect to $u$ differ from those in \eqref{POE} -- \eqref{POT}
only by a factor of $p$ from the chain rule. It cancels in the conditions $E'''(u)=0$ and $E''(u)=0$, leaving them unchanged, but not in the extremum requirement $E'(u)=0$, which becomes (cf. \eqref{POE})
\be\label{ben}
x-ux-u=0\MM{implying that}x=\frac{u}{1-u}\,.
\ee
At the critical value of the chemical potential $\mu=\mu_c=(f-1)/2$, which remains the same as before, the extremum condition for the function $E(p,\mu_c;u)$ becomes
\be\label{EXU}
\e^{p(u-1/2)}=\frac u{1-u}\,.
\ee
By contrast to \eqref{EXC}, this equation contains the parameter $p$ only in the exponential,
admitting the closed-form solution in terms of a logarithm,
\be\label{EPP}
p=\frac1{u-1/2}\,\ln\frac{u}{1-u}\,,
\ee
in contrast to its counterpart \eqref{GES} involving the Lambert function $W(z)$.
In spite of an apparent singularity at $u=1/2$, the equation \eqref{EPP} yields $p(1/2)=4$, the correct critical value $p_c$.

Substituting \eqref{EXU} and \eqref{EPP} for $x$ and $p$ in $E(p,\mu_c;u)$, we obtain the function
\be\label{blin}
E_{\rm max}(\mu_c;u)=-\ln(1-u)-\frac{u^2}{2u-1}\,\ln\frac u{1-u}
\ee
giving the coexisting curve (binodal) passing through the equal-height maxima of $E(p,\mu_c;u)$ at $\bum$, symmetric with respect to the special value $\bum=\bn=1/2$ of the "density-variable" $u$, when $p>p_c$=4.
In Section \ref{SPE}, the analogous result has been given by \eqref{eMAX} in terms of two branches
of $W(z)$.

For fixed values of $p$ and all extrema implied by the condition \eqref{ben}, we obtain isothermal curves similar to those identified in Section \ref{VWL}. In the present $u$-representation, these are given by (cf. \eqref{SDS})
\be\label{TAK}
E_{\rm ext}(p,\mu;u)=-\ln(1-u)-\frac12\,p\,u^2\,.
\ee

Restricting \eqref{TAK} to the maxima of $E(p,\mu;u)$, which correspond to the stable thermodynamic states, requires introducing the constraint: these isotherms have to lie above or touch the binodal line defined in \eqref{blin}. Therefore, the equation of state that obeys the thermodynamic stability is given by the system
\be\label{ERY}
\begin{cases}\displaystyle{
E_{\rm max}(u)=-\ln(1-u)-\frac12\,p\,u^2
}\\[2mm]\displaystyle{
E_{\rm max}(u)\ge-\ln(1-u)-\frac{u^2}{2u-1}\,\ln\frac u{1-u}\,.
}
\end{cases}
\ee

Owing to the definition of the pressure $P^*$ via \eqref{eons}, we have $pP^*=E(\bum)$ (cf. \eqref{ERR}). Since the reduced density $\bn$ is identified with $\bum$ by \eqref{TT}, the same system can be symbolically written as
\be\label{ERT}
\begin{cases}
P^*=P^*(p,\bn)
\\[1mm]
P^*\ge P_{\rm bin}^*(\bn)\,.
\end{cases}
\ee
Explicitly, for the  binodal curve $P_{\rm bin}^*(\bn)$ in the ($\bn,P^*$) plane we have
\be\label{OOP}
P_{\rm bin}^*(\bn)
=\Big(\frac12-\bn\Big)\frac{\ln(1-\bn)}{\displaystyle\ln\frac{\bn}{1-\bn}}-\frac12\,\bn^2\,,
\ee
and the equation of state in physical variables $P^*,\T,\bn$ for stable thermodynamic states is given by
\be\label{ERE}
\begin{cases}
\displaystyle{P^*=-\,\T\ln(1-\bn)-\frac12\,\bn^2}\\[3mm]
\displaystyle{\T\ge\Big(\bn-\frac12\Big)\left(\ln\frac\bn{1-\bn}\right)^{-1}},
\end{cases}
\ee
which is a consequence of \eqref{ERY} and \eqref{ERT} where we switched to the temperature variable $\T=1/p$. The first line in \eqref{ERE} directly follows from dividing its counterpart in \eqref{ERY} by $p$ and substituting $p=1/\T$, while the second line combines both equations in \eqref{ERY} into a modified binodal condition in agreement with \eqref{EPP}.

\begin{figure}[htb]
\bc
\includegraphics[width=0.48\textwidth]{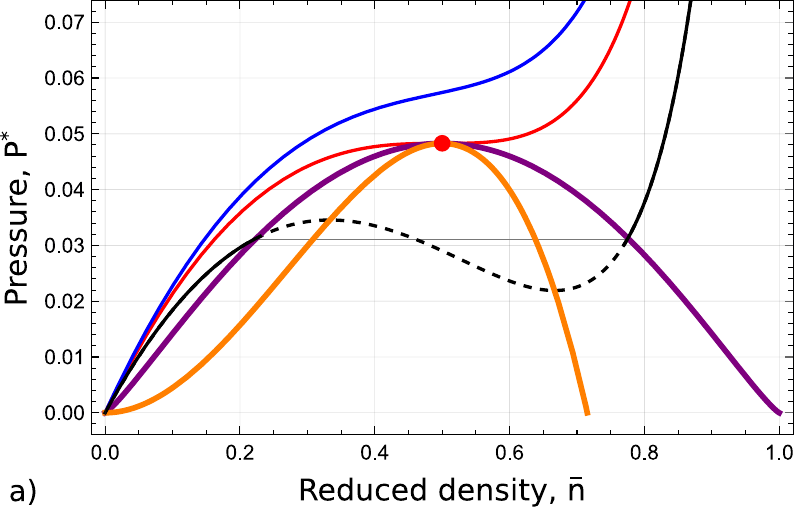}
\hfill
\includegraphics[width=0.48\textwidth]{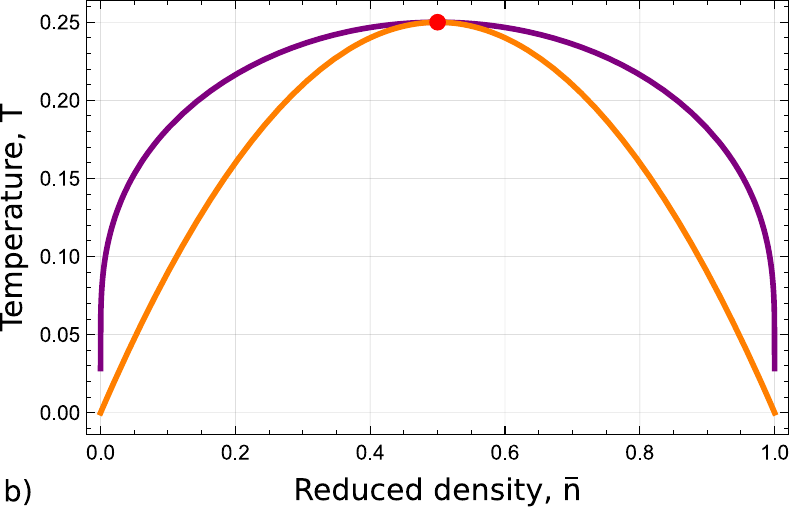}
\ec
\caption{Illustration of analytical results for the CFM in the strong-repulsion limit from Section~\ref{ph}.
    \textit{Fig.~\ref{f6}a):} The $(P^*,\bar{n})$ phase diagram showing three pressure isotherms from the equation of state \eqref{ERE} at $\T<\T_c$ (black), $\T=\T_c=1/4$ (red), and $\T>\T_c$ (blue). The subcritical isotherm forms a van der Waals loop (dashed), whose unphysical part lies below the spinodal, given by \eqref{PSP} and presented by the orange curve.
    The purple curve represents the binodal $P^*_{\rm bin}(\bar{n})$ from \eqref{OOP}.
    Only the isotherm sections above the binodal correspond to stable states; the two regions between the spinodal and binodal correspond to metastable states. The red dot marks the critical point at $\bar{n}_c=1/2$, $P^*_c=\frac14\ln2-\frac18\simeq0.048$.
    \textit{Fig.~\ref{f6}b):} The $(\T,\bar{n})$ phase diagram showing the binodal $\T_{\rm bin}(\bar{n})$ (purple) from \eqref{UUU} and the spinodal $\T_{\rm sp}(\bar{n})=\bar{n}(1-\bar{n})$ (orange) from \eqref{XAX}, both symmetric with respect to $\bar{n}=1/2$. The red dot marks the critical point at $\bar{n}_c=1/2$, $\T_c=1/4$.}
\label{f6}
\end{figure}

Physically, the first line of \eqref{ERE} represents a family of pressure isotherms as functions of the reduced density $\bn$ considered in the ($\bn,P^*$) plane, while the constraint in the second line means that the relevant temperatures $\T$ must be greater than or equal to that given by the binodal in the ($\T,\bn$) phase diagram,
\be\label{UUU}
\T_{\rm bin}(\bar{n})=
\frac{\displaystyle{\bn-\frac12}}{\displaystyle{\ln\frac\bn{1-\bn}}}\,,
\ee
as shown by the purple curve in Fig.~\ref{f6}b). The representation of the equation of state in the form \eqref{ERE} thus combines our understanding of the system's behavior from the perspectives of two different standard phase diagrams shown in Fig.~\ref{f6}.

To complete the above physical picture, we determine the spinodal curves by solving the simultaneous condition $E'(u)=E''(u)=0$, which now explicitly reads (cf. \eqref{SIX})
\be\label{SIW}
\begin{cases}
x-u(1+x)=0\\
(1+x)^2-px=0\,.
\end{cases}
\ee
For the argument of the logarithm in \eqref{SOS}, the system of equations \eqref{SIW} yields
\be\label{MOS}
1+x=(1-u)^{-1},
\ee
and for the spinodal curve in the ($\bn,\T$) plane we obtain
\be\label{XAX}
\T_{\rm sp}(\bn)=\bn(1-\bn)\,.
\ee
For the pressure spinodal in the ($P^*,\bn$) plane we derive from \eqref{SOS}, \eqref{MOS}, and \eqref{XAX}
\be\label{PSP}
P^*_{\rm sp}(\bn)=-\bn(1-\bn)\ln(1-\bn)-\frac12\,\bn^2\,.
\ee

Thus, for an extension of the equation of state to the region including metastable states, we have to change the inequality $P^*\ge P_{\rm bin}^*(\bn)$ in \eqref{ERT} to $P^*>P_{\rm sp}^*(\bn)$ where the pressure spinodal curve $P_{\rm sp}^*(\bn)$ is given by \eqref{PSP}. In the extended version of \eqref{ERE}, the second line should be replaced by the strict inequality $\T>\T_{\rm sp}(\bn)$, where $\T_{\rm sp}(\bn)$ is the spinodal temperature given by \eqref{XAX}.

\subsection{A graphical illustration and discussion}\label{PPP}

Our analytical results of the last section are illustrated in Fig.~\ref{f6}.
In the Fig.~\ref{f6}a), we show three representative isotherms $P^*(\T,\bn)$ obtained from the equation of state \eqref{ERE}, drawn at $\T<\T_c$ (black), $\T=\T_c$ (red), and $\T>\T_c$ (blue). Only their sections lying above the binodal $P_{\rm bin}^*(\bn)$ correspond to stable thermodynamic states. The spinodal curve $P_{\rm sp}^*(\bn)$ is given in orange. This curve runs through maxima and minima of van der Waals loops (dashed). The region below it would correspond to completely unstable, thermodynamically forbidden states. Two regions between $P_{\rm sp}^*(\bn)$ and $P_{\rm bin}^*(\bn)$, separated by the critical point, correspond to metastable states. The picture is in full qualitative agreement with
\cite[Fig.~5.12]{HansenMcDonald13}, where an "augmented" version of the van der Waals theory has been used with the equation of state \cite[(5.7.4)]{HansenMcDonald13}, which clearly differs from our approach.

Figure~\ref{f6}b) shows the ($\T,\bn$) phase diagram, analogous to that in Fig.~1~b). As before, the region of thermodynamically stable states is that above the purple binodal curve, which corresponds to the requirement of the second line in \eqref{ERE}. The orange curve represents the spinodal curve in the ($\bn,\T$) plane.

The pressure isotherms in the first line of the equation of state \eqref{ERE} completely agree with those in \cite[(4.61),~(4.64)]{FVbook} and \cite[(27)]{PS15} for the van der Waals lattice gas with Kac-normalized interactions. Moreover, our Fig.~\ref{f6}a) is a direct counterpart of Fig.~1 in \cite{PS15}, up to the numerical scale of the pressure axis. The difference stems from the temperature normalization. In \cite{PS15}, the choice $T_c=1$ leads to the critical pressure value of $\ln2-\frac12\simeq0.193147$ (as in \eqref{MVV}), while in our current formulation we have $\T_c=1/4$, which results in a critical pressure $P^*_c$ exactly four times smaller, approximately $0.048$.

All our plots in Fig.~\ref{f6} are drawn using \emph{explicit analytical expressions} derived in Section \ref{ph}, while in \cite[Fig.~1]{PS15}, only the pressure isotherms have been explicitly known, and the Maxwell construction has been additionally employed in producing the figure.

Finally, let us return to our extremum condition at the critical chemical potential \eqref{EPP} and its counterpart given by the second equation in \eqref{ERE}, which define the binodal curve in the (density - pressure) plane. Owing to its symmetry with respect to $u=1/2$, we introduce a positive deviation $\delta$ from this symmetry point via $u:=1/2+\delta$, which implies $1-u=1/2-\delta$. In terms of this parameter, the equation \eqref{EPP} reads
\be\label{KLK}
p\,\delta=\ln\frac{1/2+\delta}{1/2-\delta}\,.
\ee
Introducing the reduced densities $\rho_1\equiv\bn_1$ and $\rho_2\equiv\bn_2$ via
\be
\rho_1=\frac12-\delta\MM{and}\rho_2=\frac12+\delta
\ee
we obtain from \eqref{KLK} a beautiful equation
\be
\ln\rho_2-\ln\rho_1=\frac12\,p\,(\rho_2-\rho_1)\,.
\ee
The values $\rho_1$ and $\rho_2$ are interpreted as densities of coexisting phases at inverse temperatures $p>p_c$. Their difference $\Delta=\rho_2-\rho_1$ defines the order parameter appropriate for the critical point occurring at $p=p_c=4$ and $P^*=P^*_c=\frac14\,\ln2-\frac18\simeq0.048$.

Using the order parameter $\Delta$, we rewrite the equation \eqref{KLK} as
\be
2\,\frac p{p_c}\,\Delta=\ln\frac{1+\Delta}{1-\Delta}\,.
\ee
Exponentiating this equation and solving it for $\Delta$ appearing in the argument of the logarithm, we obtain it in the form
\be
\Delta=\tanh\Big(\frac{p}{p_c}\,\Delta\Big),
\ee
which agrees with \cite[(26)]{PS15}.
This result demonstrates that the strong-repulsion limit of the CFM recovers the standard Curie-Weiss equation in the absence of an external ordering field, as can be found, for example, in \cite[(12.47)]{Salinas}.

\section{Summary and outlook}

In this paper, we have presented analytical results for the cell fluid
model with Curie-Weiss interactions (CFM) in several limiting and special cases,
complementing the extensive numerical studies of earlier works
\cite{KKD20,KKD18,KD22,Retal25,DKPP26}.

Our main results are the following.

We have established the convergence of the integral representation for the CFM's grand-canonical partition function at the boundary of thermodynamic stability, $J_1=J_2$, by exploiting the asymptotic expansion of the deformed exponential function $K(r;z)$ at $z\to\infty$.
This extends the validity range of the model beyond the strict inequality
$J_2>J_1$ used in previous works, and is supported by new numerical results and
phase diagrams presented for $f=J_2/J_1=1$ in Table~\ref{tab1} and Fig.~\ref{f1}.
As a byproduct, the ideal-gas limit $J_1=J_2=0$ becomes formally
legitimate, and the well-known classical result for the ideal-gas grand
partition function is recovered in full agreement with \cite{Hill56}.

In the strong-repulsion limit $J_2\gg J_1$, we have derived
explicit analytical expressions for the critical point parameters, the
equation of state, and the binodal and spinodal curves in both the
$(\bn,P^*)$ and $(\bn,\T)$ planes. The critical point is located at
$p_c=4$, $\bar n_c=1/2$, and $P^*_c=\frac14\ln2-\frac18$, in good
agreement with previous numerical findings over a wide range of the parameter
$f\gtrsim3$. The Landau expansion around the critical point is consistent with
that of the classical lattice gas and Ising-like systems within
the mean-field approximation. The equation of state \eqref{ERE} is shown to be in full
agreement with the van der Waals lattice gas results of \cite{PS15,FVbook},
and the order parameter equation recovers the standard Curie-Weiss form.
%A notable feature of the strong-repulsion limit is the exact symmetry of the spinodal and binodal curves with respect to the critical density $\bar n_c = 1/2$ and the critical chemical potential $\mu_c=(f-1)/2$.

Several directions for future work suggest themselves. The asymptotic
analysis of $K(r;z)$ employed here retains only the steady
non-oscillating component. The complete asymptotic formula involving
the Jacobi theta function $\vartheta_3$, to appear in \cite{GS25},
will allow one to obtain explicit expressions for the parameters of
the full sequence of critical points, extending the numerical data
of Table~\ref{tab1}. It would also be of interest to investigate
whether the binodal symmetry found in the strong-repulsion limit
persists in the full model at finite values of $f$. Finally, the
connections between the CFM and multiple-occupancy quantum systems
noted in \cite{RDKPS26} deserve further analytical exploration in
the light of the present results.

\section*{Acknowledgements}
M.A.S.\ is grateful to Stefan Gerhold (TU Wien) for bringing to our
attention the connection between the function $K(r;z)$ and the deformed
exponential function of Scott and Sokal \cite{SS05}, which arose in his
communication with Friedrich Hubalek and was overlooked in \cite{DSh24}.
M.A.S.\ acknowledges the helpful assistance of Claude Sonnet 4.6 by Anthropic and Gemini by Google in final editing. The authors are deeply grateful to all warriors of the Ukrainian Armed Forces, living and fallen, for making it possible to continue the research work.
The work was supported by the National Research Foundation of Ukraine,
project No.\ 2023.03/0201.

\bibliographystyle{JHEPm}\bibliography{bankTOTAL}

@article{triple,
title = {Triple point in a cell fluid model with effective temperature-dependent
         attraction},
author = {M. P. Kozlovskii and O. A. Dobush and I. V. Pylyuk and R. V. Romanik
          and M. A. Shpot},
fjournal = {Physica A: Statistical Mechanics and its Applications},
journal = {Physica A},
volume = {692},
pages = {131510},
year = {2026},
doi = {https://doi.org/10.1016/j.physa.2026.131510}
}

@book{Salinas,
  title={Introduction to Statistical Methods},
  author={S. R. A. Salinas},
  booktitle={Introduction to Statistical Physics},
  year={2001},
  publisher={Springer},
  ADDRESS = {New York},
  doi={10.1007/978-1-4757-3508-6}
}

@BOOK{FVbook,
    AUTHOR  = {S. Friedli and Y. Velenik},
    TITLE   = {Statistical Mechanics of Lattice Systems.
               {A} Concrete Mathematical Introduction},
    YEAR    = {2018},
    PUBLISHER = {Cambridge University Press},
    ADDRESS = {Cambridge, New York, Port Melbourne, New Delhi},
    doi={10.1017/9781316882603}
}

@article{PS15,
    author = {D. Pirjol and C. Schat},
    title = {Thermodynamics of a lattice gas with linear attractive potential},
    fjournal = {Journal of Mathematical Physics},
    journal = {J. Math. Phys.},
    year = {2015},
    volume = {56},
    number = {1},
    pages = {013303},
    doi = {10.1063/1.4904833}
}

@article{NL11,
    author  = {T. Neuhaus and C. N. Likos},
    title   = {Phonon dispersions of cluster crystals},
    journal = {J. Phys.: Condens. Matter},
    fjournal = {Journal of Physics: Condensed Matter},
    year    = {2011},
    volume  = {23},
    number  = {23},
    pages   = {234112},
    doi     = {10.1088/0953-8984/23/23/234112}
}

@article{MMC23,
    author  = {M. de Mello and R. D\'{\i}az-M\'endez and A. Mendoza-Coto},
    title   = {Ultrasoft Classical Systems at Zero Temperature},
    journal = {Entropy},
    year    = {2023},
    volume  = {25},
    number  = {2},
    pages   = {356},
    doi     = {10.3390/e25020356}
}

@article{Pre14,
    author  = {S. Prestipino},
    title   = {Cluster phases of penetrable rods on a line},
    journal = {Phys. Rev. E},
    fjournal = {Physical Review E},
    year    = {2014},
    volume  = {90},
    pages   = {042306},
    doi     = {10.1103/PhysRevE.90.042306}
}

@article{WS13,
	author = {N. B. Wilding and P. Sollich},
	title = {A {M}onte {C}arlo method for chemical potential determination in single
             and multiple occupancy crystals},
	journal = {Europhys. Lett.},
	fjournal = {Europhysics Letters},
	year = {2013},
	volume = {101},
	number = {1},
	pages = {10004},
	doi = {10.1209/0295-5075/101/10004}
}

@article{PGT15,
    author  = {S. Prestipino and D. Gazzillo and N. Tasinato},
    title   = {Probing the existence of phase transitions in one-dimensional
               fluids of penetrable particles},
    journal = {Phys. Rev. E},
    fjournal = {Physical Review E},
    year    = {2015},
    volume  = {92},
    pages   = {022138},
    doi     = {10.1103/PhysRevE.92.022138}
}

@article{RDKPS26,
   title =     {Entropy of the cell fluid model with {Curie--Weiss} interaction},
   author =    {R. V. Romanik and O. A. Dobush and M. P. Kozlovskii and I. V. Pylyuk and M. A. Shpot},
   year = {2026},
   fjournal = {The European Physical Journal B},
   journal = {Eur. Phys. J. B},
   volume = {99},
   number = {1},
   pages = {13},
   doi = {10.1140/epjb/s10051-026-01119-0}
}

@article{Retal25,
	title={A multiple occupancy cell fluid model with competing attraction and repulsion interactions},
	author={Romanik, R. V. and Dobush, O. A. and Kozlovskii, M. P. and Pylyuk, I. V. and Shpot, M. A.},
	year={2026},
	fjournal = {Journal of Statistical Physics},
	journal = {J. Stat. Phys.},
	volume = {193},
	number = {2},
	pages = {14},
	doi = {10.1007/s10955-026-03568-4}
}

@article{Petal25,
  author={I. V. Pylyuk and O. A. Dobush and M. P. Kozlovskii and R. V. Romanik and M. A. Shpot},
  title={Influence of attractive parts of interaction potentials on critical point parameters},
  fjournal = {Ukrainian Journal of Physic},
  journal = {Ukr. J. Phys.},
  year={2025},
  volume = {70},
  issue = {11},
  pages ={787 -- 793},
  doi = {0.15407/ujpe70.11.787}
}

@misc{GS25,
    author={S. Gerhold and M. A. Shpot},
    title={Asymptotics of a deformed exponential function via discrete {L}aplace's method,
          to be published},
    year={2026},
    note={to be published}
}

@article{DSh24,
  author={O. A. Dobush and M. A. Shpot},
  title={{A new special function related to a discrete Gauss-Poisson distribution
  and some physics of the cell model with Curie-Weiss interactions}},
  year={2024},
  eprint={2412.05428},
  eprinttype={arXiv},
  primaryClass={math.CA},
  doi = {10.48550/arXiv.2412.05428}
}

@article{Kastner25,
  author  = {M. Kastner},
  title   = {Long-range systems, (non)extensivity, and the rescaling of energies},
  year    = {2025},
  eprint  = {2506.22296},
  eprinttype = {arXiv},
  primaryClass = {cond-mat.stat-mech},
  doi={10.48550/arXiv.2506.22296}
}

@article{R,
  title={Modular equations and approximations to $\pi$},
  author={S. Ramanujan},
  journal={Quart. J. Math.},
  volume={45},
  pages={350--372},
  year={1914}
}

@BOOK{BO78,
    AUTHOR  = {C. M. Bender and S. A. Orszag},
    TITLE   = {Advanced Mathematical Methods for Scientists and Engineers},
    YEAR    = {1978},
    PUBLISHER   = {McGraw-Hill},
    ADDRESS = {New York}
}

@book{HansenMcDonald13,
	title={Theory of Simple Liquids: with Applications to Soft Matter},
	author={J-P. Hansen and I. R. McDonald},
	edition={4th},
	isbn={9780123870339},
	lccn={2013372077},
	url={https://books.google.com.ua/books?id=BEuYUhaDvIgC},
	year={2013},
	publisher={Academic Press},
	taken_from={https://books.google.com.ua/books/about/Theory_of_Simple_Liquids.html?id=BEuYUhaDvIgC&redir_esc=y}
}

@book{Hill56,
	title={Statistical Mechanics},
	titlef={Statistical Mechanics. Principles and Selected Applications},
	author={T. L. Hill},
	year={1956},
	publisher={McGraw-Hill},
	address={New York}
}

@incollection{Fedoryuk89,
	title = {Asymptotic Methods in Analysis},
	author = {M. V. Fedoryuk},
	editor = {R. V. Gamkrelidze},
	booktitle = {Analysis {I}. Integral Representations and Asymptotis Methods},
	series = {Encyclopedia of Mathematical Sciences},
	volume = {13},
	pages = {83 -- 191},
	year = {1989},
	publisher = {Springer},
	doi = {10.1007/978-3-642-61310-4}
}

@article{KK16,
	author={{\relax Yu} Kozitsky and M. Kozlovskii},
	title = {{A phase transition in a continuum particle system
              with binary Curie-Weiss interactions}},
	year={2020},
  eprint={1610.01845v1},
  eprinttype={arXiv},
  primaryClass={math-ph},
  doi = {10.48550/arXiv.1610.01845}
}

@InCollection{KKD18,
	author={{\relax Yu}. V. Kozitsky and M. P. Kozlovskii and O. A. Dobush},
	editor={L. A. Bulavin and A. V. Chalyi},
	title={Phase Transitions in a Continuum {C}urie-{W}eiss System: {A} Quantitative Analysis},
	booktitle={Modern Problems of Molecular Physics},
	year={2018},
	publisher={Springer International Publishing},
	address={Cham},
	pages={229 -- 251},
	isbn={978-3-319-61109-9},
	doi={10.1007/978-3-319-61109-9_11},
	url={https://link.springer.com/chapter/10.1007/978-3-319-61109-9_11}
}

@article{KKD20,
	author={{\relax Yu} V. Kozitsky and M. P. Kozlovskii and O. A. Dobush},
	title = {{A phase transition in a Curie-Weiss system with binary interactions}},
	fjournal = {Condensed Matter Physics},
	journal = {Condens. Matter Phys.},
	volume={23},
	number={2},
	year={2020},
	pages={23502},
	doi={10.5488/CMP.23.23502},
	url={https://icmp.lviv.ua/journal/zbirnyk.102/23502/abstract.html}
}

@article{KD22,
	author = {M. P. Kozlovskii and O. A. Dobush},
	title = {Phase behavior of a cell model with {C}urie-{W}eiss interaction},
	fjournal = {Journal of Molecular Liquids},
	journal = {J. Mol. Liq.},
	volume = {353},
	pages = {118843},
	year = {2022},
	issn = {0167-7322},
	doi = {10.1016/j.molliq.2022.118843},
	url = {https://www.sciencedirect.com/science/article/pii/S0167732222003804}
}

@article{Ruelle70,
	author = {D. Ruelle},
	title = {Superstable interactions in classical statistical mechanics},
	fjournal = {Communications in Mathematical Physics},
	journal = {Commun. Math. Phys.},
	year = {1970},
	volume = {18},
	number = {2},
	pages = {127 -- 159},
	doi = {10.1007/BF01646091},
	url = {https://doi.org/10.1007/BF01646091}
}

@book{deBruijn,
	author={N. G. {de Bruijn}},
	title={Asymptotic methods in analysis},
	year={1958},
	publisher={North-Holland Publishing Co.},
	address={Amsterdam}
}

@book{FS,
	author={P. Flajolet and R. Sedgewick},
	title={Analytic Combinatorics},
	year={2009},
	PUBLISHER = {Cambridge University Press},
    ADDRESS = {Cambridge},
    doi={10.1017/CBO9780511801655}
}

@article{Paris11,
    author = {R. B. Paris},
    title = {The discrete analogue of {L}aplace's method},
    fjournal = {Computers and Mathematics with Applications},
    journal = {Comp. Math. Appl.},
    volume = {61},
    pages = {3024 -- 3034},
    year = {2011},
    doi={10.1016/j.camwa.2011.03.092}
}

@book{Wong,
	author={R. Wong},
	title={Asymptotic Approximations of Integrals},
	year={1989},
	PUBLISHER = {Academic Press},
    ADDRESS = {New York},
    doi={10.1016/C2013-0-07651-7}
}

@article{Lambert,
    author = {R. M. Corless and G. H. Gonnet and D. E. G. Hare and D. J. Jeffrey
     and D. E. Knuth},
    title = {{On the Lambert $W$ function}},
    fjournal = {Advances in Computational Mathematics},
    journal = {Adv. Comput. Math.},
    volume = {5},
    pages = {329 -- 359},
    year = {1996},
    doi={10.1007/BF02124750}
}

@article{DKPP26,
author={O. A. Dobush and M. P. Kozlovskii and I. V. Pylyuk and {\relax Yu} O. Plevachuk},
title = {Influence of microscopic parameters on phase behavior of a cell model with {C}urie-{W}eiss interaction},
	journal = {Physica A},
	fjournal = {Physica A: Statistical Mechanics and its Applications},
	volume = {681},
	pages = {131133},
	year = {2026},
	doi = {10.1016/j.physa.2025.131133}
}

@article{SS05,
  title={{The repulsive lattice gas, the independent-set polynomial, and the
         Lov{\'a}sz local lemma}},
  author={A. D. Scott and A. D. Sokal},
  fjournal={Journal of Statistical Physics},
  journal="J. Stat. Phys.",
  volume={118},
  pages={1151 -- 1261},
  year={2005},
  doi = {10.1007/s10955-004-2055-4}
}

@article{Sokal12,
author = {A. D. Sokal},
title = {The leading root of the partial theta function},
аjournal = {Advances in Mathematics},
journal = {Adv. Math.},
volume = {229},
number = {5},
pages = {2603 -- 2621},
year = {2012},
doi = {https://doi.org/10.1016/j.aim.2012.01.012}
}

@article{deB49,
    author = {N. G. de Bruijn},
    title = {The Asymptotically Periodic Behavior of the Solutions of Some
             Linear Functional Equations},
    fjournal = {American Journal of Mathematics},
    journal = {Amer. J. Math.},
    volume = {71},
    number = {2},
    pages = {313 -- 330},
    year = {1949}
}

@article{deB53,
    author = {N. G. de Bruijn},
    title = {{The difference-differential equation $F'(x)=\rm{e}^{\alpha x+\beta}F(x-1)$. I, II}},
    fjournal = {Indagationes Mathematicae (Proceedings, Series A)},
    journal = {Indag. Math. (Proceedings, Ser. A)},
    volume = {15},
    number = {5},
    pages = {449 -- 464},
    year = {1953}
}

@article{EE88,
  author =        {T. Eisele and R. S. Ellis},
  journal =       {J. Stat. Phys.},
  number =        {1/2},
  pages =         {161--202},
  title =         {Multiple phase transitions in the generalized
                   {C}urie-{W}eiss model},
  volume =        {52},
  year =          {1988},
  doi =           {10.1007/BF01016409}
}

@book{Campa,
  author    = {A. Campa and T. Dauxois and D. Fanelli and S. Ruffo},
  title     = {Physics of Long-Range Interacting Systems},
  publisher = {Oxford University Press},
  year      = {2014},
  address   = {Oxford}
}

@article{Likos2001,
   author  = {C. N. Likos},
  title   = {Effective interactions in soft condensed matter physics},
  journal = {Phys. Rep.},
  volume  = {348},
  pages   = {267},
  year    = {2001},
  doi     = {10.1016/S0370-1573(00)00141-1}
}
\addcontentsline{toc}{section}{References}

\end{document}